\documentclass[fleqn,aps,prm,superscriptaddress,preprint,nobibnotes,]{revtex4-2}
\usepackage{color}
\usepackage{graphicx}
\usepackage{dcolumn}
\usepackage{bm}
\usepackage{CJKutf8}
\usepackage{float}
\usepackage{geometry}
\usepackage{changepage}
\usepackage{makecell}
\usepackage{ulem}
\usepackage{hyperref}
\usepackage{amsmath}
\usepackage[markup=default]{changes}
\usepackage{changepage}
\usepackage{changes}

\usepackage{xcolor}%

\definecolor{ao(english)}{rgb}{0.0, 0.42, 0.24}

\begin{document}


\title[Sample title]{
A Deep Learning Potential for Accurate Shock Response Simulations in Tin
}

\author{Yixin Chen}
\affiliation{ 
HEDPS, CAPT, School of Physics and College of Engineering, Peking University, Beijing 100871, People’s Republic of China
}
\author{Xiaoyang Wang}
\affiliation{ 
Institute of Applied Physics and Computational Mathematics,
Beijing 100094, People’s Republic of China
}
\author{Wanghui Li}
\affiliation{
Institute of High Performance Computing, Agency for Science, Technology and Research, Singapore 138632, Republic of Singapore
}
\author{Mohan Chen}
\email{mohanchen@pku.edu.cn}
\affiliation{ 
HEDPS, CAPT, School of Physics and College of Engineering, Peking University, Beijing 100871, People’s Republic of China
}
\affiliation{
AI for Science Institute, Beijing 100084, People’s Republic of China
}
\author{Han Wang}
\email{wang$\textunderscore$han@iapcm.ac.cn}
\affiliation{ 
Institute of Applied Physics and Computational Mathematics,
Beijing 100094, People’s Republic of China
}

\begin{abstract}

Tin (Sn) plays a crucial role in studying the dynamic mechanical responses of ductile metals under shock loading. 
Atomistic simulations serves to unveil the nano-scale mechanisms for critical behaviors of dynamic responses.
However, existing empirical potentials for Sn often lack sufficient accuracy when applied in such simulation. 
Particularly, the 
solid-solid phase transition behavior of Sn poses significant challenges to the accuracy of interatomic potentials. 
%
%
To address these challenges, this study introduces a machine-learning potential model for Sn, specifically optimized for shock-response simulations. 
%
The model is trained using a dataset constructed through a concurrent learning framework and is designed for molecular simulations across thermodynamic conditions ranging from 0 to 100~GPa and 0 to 5000~K, encompassing both solid and liquid phases as well as structures with free surfaces. 
%
It accurately reproduces density functional theory (DFT)-derived basic properties, experimental melting curves, solid-solid phase boundaries, and shock Hugoniot results.
This demonstrates the model's potential to bridge ab initio precision with large-scale dynamic simulations of Sn.
\end{abstract}

                             
\maketitle


\section{\label{sec:level1}introduction}
\label{sec_intro}

Metals subjected to dynamic loading exhibit distinctive behaviors, which have been extensively investigated in both academic research and engineering applications~\cite{Novikov1981,RevModPhys.49.523,NatureMaterials2006}.
Among them, tin (Sn) is of particular interest as it is a typical representative of low-melting-point ductile metals, whose dynamic mechanical responses have been studied widely in experiments~\cite{de2007experimental,de2007spallation, Ashitkov, Kanel2015}.

Previous studies have revealed a complex pressure-temperature (P-T) phase diagram of Sn.
In detail, at 286~K and ambient pressure, Sn undergoes a well-known phase transition from a covalent-bonding diamond structure ($\alpha$-Sn, in $Fd\bar{3}m$ space group) to a metallic-bonding double-bct structure ($\beta$-Sn, in $I4_{1}/amd$ space group)~\cite{SMITH198569}.
As pressure increases, a variety of high-pressure phases have been identified, including  bct ($I4/mmm$), bcc ($Im\bar{3}m$), hcp ($P6_{3}/mmc$), and a bco ($Immm$) phase whose presence is still under debate~\cite{Liu1986CompressionsAP, olijnyk1984phase, salamat2011dense}.
Moreover, when subjected to shock loading, two phase transitions of Sn have been observed: a $\beta$-bct solid phase transition occurring at 9-10 GPa~\cite{Brigs2019Observation} and a solid-liquid phase at nearly 39 GPa~\cite{hu2008shock}. 

Beyond phase transitions under high pressure, the dynamic fracture behavior of Sn during shock loading, specifically spallation, is of significant interest to the shock physics community.
The shock induced spallation has been experimentally characterized as classical-spall and micro-spall.
The former is usually featured by the production of solid-state spall layers, while the later is generally characterized by dispersed liquid debris and extensive cavitation~\cite{Signor2009}. 
The transition pressure between classic and micro spall has been reported to be at approximately 40 GPa experimentally using optical microscopy technique~\cite{de2007experimental,de2007spallation}. 
Shock experiments can also estimate the spall strength via the measurement of free surface velocities~\cite{Ashitkov}. 
The spall strength of Sn depends on the state of materials at the time of fracture.
It has been reported that spall strength in classic spall is usually an order of magnitude higher than that in the micro-spall~\cite{Kanel2015}.
Moreover, the spall strength is sensitive to the strain rate.
The spall strength of liquid Sn increases from 0.1~GPa at $10^5$ strain rate, to 1.9~GPa at $10^9$ strain rate~\cite{Ashitkov}.

Despite the wealth of experimental results, substantial challenges remain in fully understanding the mechanisms of the phase transitions and dynamic response of Sn which requires high-resolution diagnostic techniques. 
A primary challenge is the difficulty of observing ultra-fast dynamic behaviors in situ during experiments.
Phenomena such as solid phase transitions and the nucleation of microscopic voids and cavities often originate from material behaviors occurring at atomic or lattice spatial scales and occur within picosecond or nanosecond timeframes~\cite{kadau2002microscopic,doi:10.1126/sciadv.aaz5132,silvia2022NC}.

The existence of technical challenges in extreme experiments calls for advanced simulations. 
Atomic-scale simulation has emerged as a vital alternative approach, enabling direct investigation into the underlying atomisitic mechanisms for these complex material responses.
In particular, the non-equilibrium molecular dynamics (NEMD) methods have been extensively employed in studying the shock compression and spallation  of materials~\cite{science1998,10.1063/1.1483551,LI201951,PhysRevB.74.144110,ref6DandS-microjetting}.
These studies have successfully addressed numerous microscopic processes associated with mechanical responses under shock-wave loading.
For example, Liao \textit{et al.}~\cite{ref4YiLiao} revealed that the degree of melting has a substantial effect on void distributions and thermodynamic pathways in both classical spallation and micro-spallations.
Wang \textit{et al.}~\cite{ref5Xin-XinWang}  studied the dependence of spall strength on shock velocities.
Durand \textit{et al.}~\cite{ref6DandS-microjetting} carried out large-scale MD simulations and studied the impact of phase transitions on micro-jetting and ejecta production in shock-loaded Sn.
Wu \textit{et al.}~\cite{wu2018molecular} examined the properties of ejecta under supported and unsupported shocks and identified multiple phases of Sn using the adaptive common neighbor analysis (a-CNA) method~\cite{stukowski2012structure}. 
These simulations have highlighted the complex dynamic response of Sn under shock loading and also signified effect of corresponding phase transitions in Sn.

The accuracy of NEMD simulations is largely dependent on the precision of the potential energy surface (PES) model. 
All of the studies mentioned above employed empirical potentials to represent the PES of tin.
Liao \textit{et al.}, Durand \textit{et al.} adopted the modified embedded atom method (MEAM)~\cite{ref9meam} model parameterized by Ravelo \textit{et al.}~\cite{ref10meam1997}. 
Wang \textit{et al.} and Wu \textit{et al.} applied the embedded atom method (EAM) model~\cite{ref7eam} parameterized by Sapozhnikov \textit{et al.}~\cite{ref8eam2014}.
Despite the extensive application of empirical potentials in tin simulations, existing empirical models are unable to concurrently characterize both the $\beta$ and bct phases in shock-loaded simulations of Sn.
The EAM potential by Sapozhnikov \textit{et al.}~\cite{ref8eam2014} is only fitted to the bct and bcc phases at high pressures, which fails to reproduce the $\beta$ phase at ambient pressures. 
Consequently, when employing this EAM model in MD simulations, the initial structural configuration under ambient pressure conditions must be set to bct or bcc rather than the experimentally prevalent $\beta$ phase~\cite{ref6DandS-microjetting}.
The MEAM model by Ravelo \textit{et al.}~\cite{ref10meam1997} and a more recent model by Ko \textit{et al.}~\cite{ref11meam2018Ko}, are designed to capture the low-pressure $\alpha$ and $\beta$ phases, whereas the bct and bcc phases are not included as fitting targets in these models.
Consequently, these MEAM models may provide an inaccurate description of high-pressure phase transitions in dynamic loading simulations. 
The shortcomings in these models underscore the need for the development of more precise models in this area, which serves as the motivation for our research.

To this end, machine learning interatomic potentials (MLIPs), which have emerged as a powerful approach for describing the inter-atomic PES,  provide an alternative way to overcome the long-standing challenge~\cite{PhysRevLett.98.146401,PhysRevLett.104.136403,Montavon_2013,Schütt2017,PhysRevLett.120.143001}. 
Recently, two MLIPs for Sn have been proposed, i.e., DP-SCAN~\cite{ref1DP-SCAN} and EAM-R~\cite{ref12EAM-R}, both trained based on datasets labeled by density functional theory (DFT)~\cite{PhysRev.136.B864, PhysRev.140.A1133} calculations. 
%
One major difference between the two MLIPs is the choice of exchange-correlation functional (XC functional). 
DP-SCAN used Strongly Constrained and Appropriately Normed Semilocal Density Functional(SCAN)~\cite{ref24SCANfunc}, while EAM-R used Perdew-Burke-Ernzerhof (PBE) functional~\cite{PBE}. 
We emphasize that the SCAN functional is more suitable than the PBE functional for accurately describing the phase diagram of Sn. 
Models trained with data from the PBE functional have failed to reproduce the correct phase diagram of Sn \cite{ref1DP-SCAN}, as corroborated by the AIMD studies of Rehn \textit{et al.}~\cite{PhysRevB.103.184102}.
Therefore, the SCAN functional is adopted in our work. 
However, the primary limitation of these exisiting MLIPs in simulating dynamic mechanical responses lies in their insufficient dataset coverage, which lacks essential thermodynamic ranges.
{The training data of DP-SCAN are restricted to pressures below 50 GPa and temperatures below 2000 K.
While the data set of EAM-R contains volumetric strains up to 20\% and structural distortion of 5\%, which roughly correspond to 18 GPa and 1200 K, respectively.}
The range of the two MLPs cannot fully cover the thermodynamic range of interest 
relevant to dynamic loading experiments and simulations~\cite{ref13HuExp, de2014microjetting, Saunders2024, wu2018molecular,10.1063/5.0099331, 10.1063/5.0003089}. 
More importantly, they do not account for the properties of Hugoniot states (the final steady states of material after a shock wave passes by)~\cite{Needham2018}.
These limitations highlight the necessity for a new Sn potential model specifically suited for dynamic loading simulations.

In this work, we present a new MLIP for Sn that is suitable in simulating dynamic responses of Sn. 
We utilize the Deep Potential (DP) method, which has been successfully applied to Sn and to many other systems~\cite{Jiang_2021,NC2020,PhysRevLett.126.236001,PhysRevB.102.214113,Liu_2020,10.1063/5.0030123,Wang_2022,PhysRevMaterials.7.093601,D2CP04105G,POD2024}. 
The proposed model, named DP-SCAN-S, expands the training dataset of DP-SCAN through exploration within a pressure range of 0–100 GPa and a temperature range of 0–5000 K \cite{ref3dpgen}.
The DP-SCAN-S model maintains the accuracy of DP-SCAN in capturing the fundamental properties of Sn.
More importantly, it excels in accurately describing phenomena under dynamic loading conditions, such as Hugoniot curves, stacking faults, and fractures.
Additionally, we incorporate configurations relevant to free surfaces to improve characterization of dynamic fracture. 

The organization of the paper is as follows: the section II details the methods of developing the MLIP that covers a unprecedented wide range of pressure and temmperature as well as the consideration of free surfaces effects of the sampling.
The section III displays the comprehensive results and discussions including basic properties, generalized stacking fault energy, shock Hugoniot and phase diagrams as well as melting points. 
Lastly, the conclusions are given.

\section{Methods}
The dataset of DP-SCAN-S is extended based on the dataset of DP-SCAN using the concurrent learning scheme DP-GEN\cite{ref3dpgen}, an iterative workflow that runs through the  \textbf{training}, \textbf{exploration}, and \textbf{labeling}  processes. 
\subsection{Training}
The training of the DP models are implemented by the DeePMD-kit package\cite{wang2018deepmd}.
The detailed principles of the DP models are described in Ref.\cite{ref2dp}.
DP-SCAN-S applies the same model structure as DP-SCAN, thus their training protocols are similar: 
during the training, the loss function,
\begin{align}
\label{eq:loss}
    \mathcal L &= \frac{1}{\vert \mathcal B\vert} 
    \sum_{k\in \mathcal B}
    \Big(
     p_\epsilon \frac 1N \vert \hat E^k - E^k\vert^2 + 
     p_f \frac 1{3N} \sum_{i\alpha} \vert \hat F_{i\alpha}^k - F_{i\alpha}^k \vert^2 +
     p_\xi \frac 1{9N} \sum_{i\alpha} \vert \hat \Xi_{\alpha\beta}^k - \Xi_{\alpha\beta}^k \vert^2
    \Big), 
\end{align}
is minimized.
$\mathcal{B}$ is a mini-batch of training data.
$\vert\hat E^k - E^k\vert$, $\vert\hat F_{i\alpha}^k - F_{i\alpha}^k\vert$, and $\vert\hat \Xi_{\alpha\beta}^k - \Xi_{\alpha\beta}^k\vert$ represent the differences between model predicted energies ($E$), forces ($F$) and virial tensors ($\Xi$) and those in the DFT labels.
The relative weight of energy, forces, and virial tensor to the loss function are determined by ($p_\epsilon$, $p_f$, $p_\xi$), a set of prefacters which are adjusted during training. 
At the beginning of the training, we set the prefactors ($p_\epsilon,p_f,p_\xi$) to (0.02, 1000, 0.02) and the learning rate to $10^{-3}$. 
($p_\epsilon,p_f,p_\xi$) at the end of training are set to (2.0, 1.0, 2.0), and the learning rate decays to $1\times10^{-8}$. 
During the training step in each iteration, an ensemble of four models are trained with the same dataset and hyper-parameters, but with different random seeds for parameter initialization. 

\subsection{Exploration}
The configurational space is explored using Deep Potential molecular dynamics (DPMD)\cite{PhysRevLett.120.143001} simulations, performed using the LAMMPS package\cite{ref18lammps}. 
During the exploration, the MD trajectory is analyzed by calculating the maximum force deviation of frames along the trajectory using the four models. 
\begin{equation}
\epsilon_{t}=\max_{i}\sqrt{\langle \| \bm{F}_{\omega,i}- \langle \bm{F}_{\omega,i} \rangle \| ^{2} \rangle},
\end{equation}
where \(\bm{F}_{\omega,i}\) denotes the force acting on atom \(i\) as predicted by the model.
When the model deviation exceeds the upper bound, \(\sigma_{\mathrm{high}}\), the frame is considered nonphysical and is excluded from DFT labeling.
When the model deviation is below the lower bound, \(\sigma_{\mathrm{low}}\), the force prediction accuracy is deemed satisfactory.
Configurations with force deviation within the range \(\sigma_{\mathrm{low}} \leq \epsilon_t < \sigma_{\mathrm{high}}\) are considered as candidates for subsequent DFT labeling. 
With the progress of the iterations, the ratio of candidate frames whose $\epsilon_t< \sigma_{\mathrm{low}}$ gradually converges. 
The iteration are stopped when 99.5\% of the explored frames exhibit \(\epsilon_t< \sigma_{\mathrm{low}} \).

The exploration consists of two stages.
In the first stage, we extend the pressure-temperature $(P,T)$ range of the bulk structures in the DP-SCAN dataset to 0-5,000 K and 0-100 GPa.
The $\alpha$, $\beta$, $bct$ and $bcc$ structures are used initial configurations.  $\alpha$ phase is excluded in exploration with  $P >50$~GPa due to its instability in the high-pressure regime.
The NPT ensemble is applied in explorations of these bulk structures.
During the exploration steps, $\sigma_{\mathrm{high}}$ was set to 0.3~eV/$\mathrm{\AA}$, while different $\sigma_{\mathrm{low}}$ were chosen to balance the cost and the abundance of the data. 
Specifically, under $0<P\leq50$ GPa and $0<T\leq2,000$ K, $\sigma_{\mathrm{low}}$ was set to 0.12~eV/$\mathrm{\AA}$, which was the same as that of the DP-SCAN model. 
In the range of  $50< P\leq100$ GPa and $0<T\leq2,000$ K, $\sigma_{\mathrm{low}}$ was chosen as 0.05~eV/$\mathrm{\AA}$. 
In the range of $0<P\leq100$ GPa and $2000<T\leq5,000$ K, $\sigma_{\mathrm{low}}$ was set to 0.09~eV/$\mathrm{\AA}$.

In the second stage, the free-surface structure of \(\beta\)-Sn and bct-Sn are explored.
The \(\beta\)-Sn and bct-Sn crystal structures were first relaxed with DFT under pressures ranging from 0 to 30 GPa and 0 to 50 GPa, respectively. 
Using the relaxed structures, the (100) and (001) free surfaces for the two phases are created by separating the bulk structures by 0.5, 1, 2, 4, 6, 8, and 12 Å, which represents the decohesion process during the dynamic fracture.
During the exploration of the surface structures, an NVT ensemble was applied with temperature ranging from 0-2000~K and $\sigma_{low}$ set to 0.09~eV/$\mathrm{\AA}$.
In this stage, virial tensors of the surface structures are not the fitting target during training, so we set the training prefactor $p_\xi$ to 0.

\subsection{Labeling.}
\label{sec:dft}
The DFT calculations were performed with the Vienna ab-initio simulation package (VASP 5.4.4)~\cite{ref23VASP}. 
We used the SCAN XC functional~\cite{ref24SCANfunc} as discussed in Sec.~\ref{sec_intro}. 
We adopted the same DFT parameters as DP-SCAN\cite{ref1DP-SCAN} to keep the consistency of the data set. 
The energy cutoff for the plane-wave basis set is set to 650~eV.
The gaussian smearing method with a smearing width of 0.2~eV was applied.
The spacing of the Monkhorst-pack (MP) $k$-point sampling grid~\cite{ref25kpointsM&P} in the momentum space was 0.10~$\mathrm{\AA}^{-1}$.
The convergence threshold in self-consistent electronic iterations was set to $10^{-6}$~eV.

\subsection{Productive training and refinement.}
In total, the dataset contains 15,309 configurations, in which 6,647 are inherited from DP-SCAN and the rest are collected in this work.
The productive DP-SCAN-S model is finally trained with long steps.
Four models were trained with 16,000,000 steps, with ($p_\epsilon$, $p_f$, $p_\xi$) set to (0.02, 1000, 0.02) at the beginning of the training and (1.00, 1.00, 1.00) at the end of the training. The learning rate at the begining of the training is $1\times10^{-3}$.
The model with the least training error is further refined with additional 16,000,000 training steps. During the refinement, the model parameters are initiated from the productive model, rather than generated randomly.  The ($p_\epsilon$, $p_f$, $p_\xi$) are set to (10, 1.00, 0.02) at the beginning of the refinment and (10, 1.0, 0.5) at the end of the refinement.

The training root mean square error (RMSE) of DP-SCAN-S on total energy is 7.55 meV/atom, the RMSE of atomic force is 9.35$\times10^{-2}$ eV/$\mathrm{\AA}$, and the RMSE of virial is 4.51$\times10^{-2}$ eV/atom.

\section{\label{sec:level1}Results and DISCUSSIONS}

\subsection{Basic properties}

\begin{table}
\footnotesize
\centering
\caption{Properties of $\beta$-Sn and bct-Sn at the ground state, which are computed by the DFT method with SCAN functional (DFT-SCAN), the MEAM (Ko)~\cite{ref11meam2018Ko} model, the ML potential EAM-R~\cite{ref12EAM-R}, the DP-SCAN~\cite{ref1DP-SCAN} and DP-SCAN-S models.
Available experimental results under finite temperature and pressure are also shown.
The listed properties include the lattice parameter $a$, $c$ ($\mathrm{\AA}$), the $c/a$ ratio, the elastic constants $C_{11}$, $C_{12}$, $C_{13}$, $C_{33}$, $C_{44}$, $C_{66}$ (GPa), the cohesive energy $E_c$ (eV), the vacancy energy $E_{\mathrm{vac}}$ (eV), and the surface energies for (001) $E_{\mathrm{surf}}^{(001)}$ and (100) $E_{\mathrm{surf}}^{(100)}$ (erg/cm$^2$) surfaces of the $\beta-$Sn structure.}
\label{table:bulk}
\setlength{\tabcolsep}{8 pt}
\scriptsize
\begin{tabular}{lllllll}
\hline
\hline
$\beta$-Sn & DFT-SCAN & DP-SCAN-S & DP-SCAN~\cite{ref1DP-SCAN} & EAM-R~\cite{ref12EAM-R} & MEAM (Ko)~\cite{ref11meam2018Ko} & Experiments\\
\hline
$a$ & 5.909 & 5.890 & 5.894~\cite{ref1DP-SCAN} & 5.924 & 5.859 & 5.831~\cite{ref28Exp3}\\
$c$ & 3.164 & 3.169 & 3.172~\cite{ref1DP-SCAN} & 3.245 & 3.206 & 3.184~\cite{ref28Exp3}\\
$c/a$ & 0.536 & 0.538 & 0.538~\cite{ref1DP-SCAN} & 0.548 & 0.547 & 0.546~\cite{ref28Exp3}\\
$C_{11}$ & 108.6 & 95.3 & 94.6~\cite{ref1DP-SCAN} & 59.1 & 89.7 & 73.4~\cite{ref28Exp3}\\
$C_{12}$ & 11.5 & 34.1 & 36.2~\cite{ref1DP-SCAN} & 53.2 & 46.7 & 59.9~\cite{ref28Exp3}\\
$C_{13}$ & 33.6 & 25.6 & 26.0~\cite{ref1DP-SCAN} & 30.7 & 36.9 & 39.1~\cite{ref28Exp3}\\
$C_{33}$ & 103.4 & 110.4 & 89.6~\cite{ref1DP-SCAN} & 73.1 & 93.7 & 90.7~\cite{ref28Exp3}\\
$C_{44}$ & 24.4 & 17.9 & 24.1~\cite{ref1DP-SCAN} & 15.6 & 7.9 & 22.0~\cite{ref28Exp3}\\
$C_{66}$ & 24.7 & 26.1 & 29.7~\cite{ref1DP-SCAN} & 19.4 & 10.6 & 23.9~\cite{ref28Exp3} \\
$E_c$ & -3.385 & -3.388 & -3.397 & -3.094 & -3.102 & -3.10~\cite{ref26Exp1}\\
$E_{\mathrm{vac}}$ & 0.786 & 0.812 & 0.900 &  0.741~\cite{ref12EAM-R} & 0.857 & \\
$E_{\mathrm{surf}}^{(001)}$ & 429.7
 & 447.1 & 432.5 & 409.622~\cite{ref12EAM-R} & 393~\cite{ref11meam2018Ko} & \\
$E_{\mathrm{surf}}^{(100)}$ & 463.5
 & 467.0 & 440.6 & 294.673~\cite{ref12EAM-R} & 345~\cite{ref11meam2018Ko} & \\
\hline
\hline
bct-Sn & DFT-SCAN & DP-SCAN-S & DP-SCAN~\cite{ref1DP-SCAN} & EAM-R~\cite{ref12EAM-R} & MEAM (Ko)~\cite{ref11meam2018Ko} & Exp.\\
\hline
$a$ & 4.016 & 4.009 & 4.008~\cite{ref1DP-SCAN} & 4.040 & 3.906 & 3.70~\cite{barnett1966x}\\
$c$ & 3.361 & 3.362 & 3.365~\cite{ref1DP-SCAN} & 3.413 & 3.397 & 3.37~\cite{barnett1966x}\\
$c/a$ & 0.837 &  0.838 &  0.840~\cite{ref1DP-SCAN} &  0.845 & 0.869 & 0.91~\cite{barnett1966x}\\
$C_{11}$ & 61.4 & 78.0 & 57.3~\cite{ref1DP-SCAN} & 46.2 & 89.8 & \\
$C_{12}$ & 37.3 & 32.6 & 34.9~\cite{ref1DP-SCAN} & 42.5 & 64.8 & \\
$C_{13}$ & 49.1 &  43.2 & 38.5~\cite{ref1DP-SCAN} & 47.5 & 49.5 & \\
$C_{33}$ & 84.6 & 67.3 & 74.3~\cite{ref1DP-SCAN} & 69.3 & 73.6 & \\
$C_{44}$ & 11.3 & 14.1 & 11.3~\cite{ref1DP-SCAN} & 14.7 & 42.5 & \\
$C_{66}$ & 26.9 & 34.3 & 23.7~\cite{ref1DP-SCAN} & 31.7 & 52.4 & \\
$E_c$ & -3.278 & -3.281 & -3.290 & -3.078 & -3.082 & \\
\hline
\hline
\end{tabular}
\end{table}

TABLE~\ref{table:bulk} presents the basic properties of $\beta$-Sn and bct-Sn at the ground state.
We compare the properties computed by DFT method with the SCAN functional (DFT-SCAN), and those predicted by various potentials, i.e., the MEAM potential parameterized by Ko \textit{et al.}~\cite{ref11meam2018Ko}, EAM-R~\cite{ref12EAM-R}, DP-SCAN~\cite{ref1DP-SCAN} and DP-SCAN-S. 
The reference experimental value is also listed for comparisons.
The basic properties include the lattice parameters \(a\) and \(c\), the \(c/a\) ratio, elastic constants \(C_{11}\), \(C_{12}\), \(C_{13}\), \(C_{33}\), \(C_{44}\), \(C_{66}\), cohesive energy \(E_c\), vacancy formation energy \(E_{\mathrm{vac}}\), and the surface formation energies for the (001) and (100) surfaces, denoted as \(E_{\mathrm{surf}}^{(001)}\) and \(E_{\mathrm{surf}}^{(100)}\).

As expected, the DP-SCAN-S model demonstrates similar accuracy to the DP-SCAN model for both \(\beta\) and bct phases of Sn in terms of the basic material properties, since the DP-SCAN-S training dataset is constructed based on that of DP-SCAN. 
The two DP models thus yield close predictions for the lattice parameters and elastic constants \(C_{11}\), \(C_{12}\), and \(C_{13}\) for the \(\beta\)-Sn structure. 
Discrepancies exist between the DP models and DFT results.
For instance, DP-SCAN-S and DP-SCAN predict $C_{12}$ of $\beta$ phase as 34.1 and 36.2 GPa, respectively, substantially larger than the 11.5 GPa calculated by DFT-SCAN. 
Additionally, the \(C_{44}\) value predicted by DP-SCAN-S is 17.9 GPa, which is lower than the DFT result of 24.7 GPa. 
DP-SCAN-S is more accurate than DP-SCAN on the prediction of $C_{33}$ and $C_{66}$ in reference  to DFT value, but less accurate on $C_{44}$.
In comparison with DP-SCAN, DP-SCAN-S significantly improves the prediction of the cohesive energy $E_c$ of $\beta$-Sn by reducing the deviation from 12~meV/atom (observed with DP-SCAN) to 3~meV/atom compared to DFT results. 

DP-SCAN-S predicts the vacancy formation energy  to be 0.812 eV, a better agreement with DFT result than DP-SCAN, EAM-R and the MEAM(Ko).
DP-SCAN-S and DP-SCAN have similar accuracy on the prediction of the formation energy of (001) and (100) free surfaces of $\beta$-Sn. 
DP-SCAN is more accurate at the (001) surface, while DP-SCAN-S is more accurate at (100) free surface.
In reference with the DFT results, where the (100) surface have high formation energy than (001) surface, EAM-R and MEAM(Ko) are qualitatively incorrect on the relative formation energy of the two free surfaces.

For the bct-Sn structure, DP-SCAN-S improves the predictions of cohesive energies and lattice parameters, in comparison with DP-SCAN. 
However, except for \(C_{13}\), DP-SCAN-S yields less accuracy for elastic constants. This is likely attributed to the incorporation of additional high-pressure data, mainly consisting of bct configurations, into the DP-SCAN-S training set. 
Such an inclusion adds complexity to the training process, thus undermining the accuracy of elastic constants calculated at 0 GPa.

In addition, DP-SCAN-S is more accurate than the MEAM(Ko) potential on the prediction of cohesive energy and elastic constants, especially for the $bct$ structures, since the elastic constants for $bct$-Sn is not the fitting target of MEAM(Ko). 
%
%
%
%
The discrepancies between DP-SCAN-S and EAM-R lie in the different choices of XC functional on DFT training data, which has been discussed in Sec.~\ref{sec_intro}.
%
%
We found that the relative error of the lattice constant $a$ of $\beta$-Sn predicted by DP-SCAN-S, compared with the experimental data~\cite{ref28Exp3}, is 1\%, while that of $c$ is 0.5\%. For EAM-R, these two relative errors are 1.6\% and 1.9\%, respectively.
However, the $c/a$ ratio obtained by DFT-SCAN is 1.8\% lower than the experimental value~\cite{ref28Exp3}, and the discrepancy is also reflected in the results of DP-SCAN-S. 

\subsection{Generalized Stacking Fault Energy} 


The generalized stacking fault energy (GFSE) reflects the energy changes associated with rigid displacement of a part of crystal against another part along a given slip plane.
It is a critical concept in materials science which provides insights into the capability of plastic deformation of materials. 
Specifically, the unstable fault energy (USFE) corresponds to the peak on the GFSE profile, representing the energy barrier to be overcome for a stacking fault to form and propagate. 
This energy barrier is crucial for understanding the stability and mobility of defects in materials, as it determines the difficulty of fault formation and the material's resistance to plastic deformation.

\begin{figure}[H]
    \centering
    \includegraphics[width=.5\textwidth]{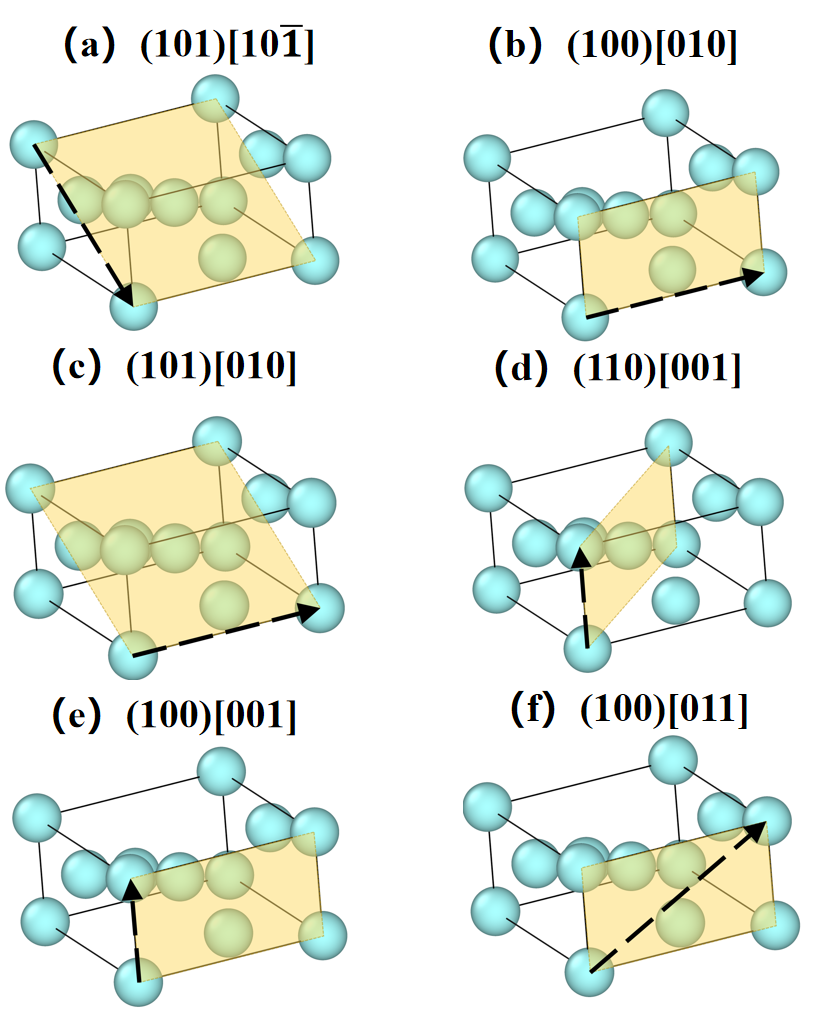}
    \caption{Illustrations of the 6 calculated slip systems of Sn, including (a) (101)[10$\overline{1}$]; (b) (100)[010]; (c) (101)[010]; (d) (110)[001]; (e) (100) [001];  (f) (100)[011]. Yellow planes denote the crystal planes where stacking faults occur, and black arrows indicate the slip direction. }
    \label{fig:SYS}
\end{figure}

\begin{figure}[H]
    \centering
    \includegraphics[width=.9\textwidth]{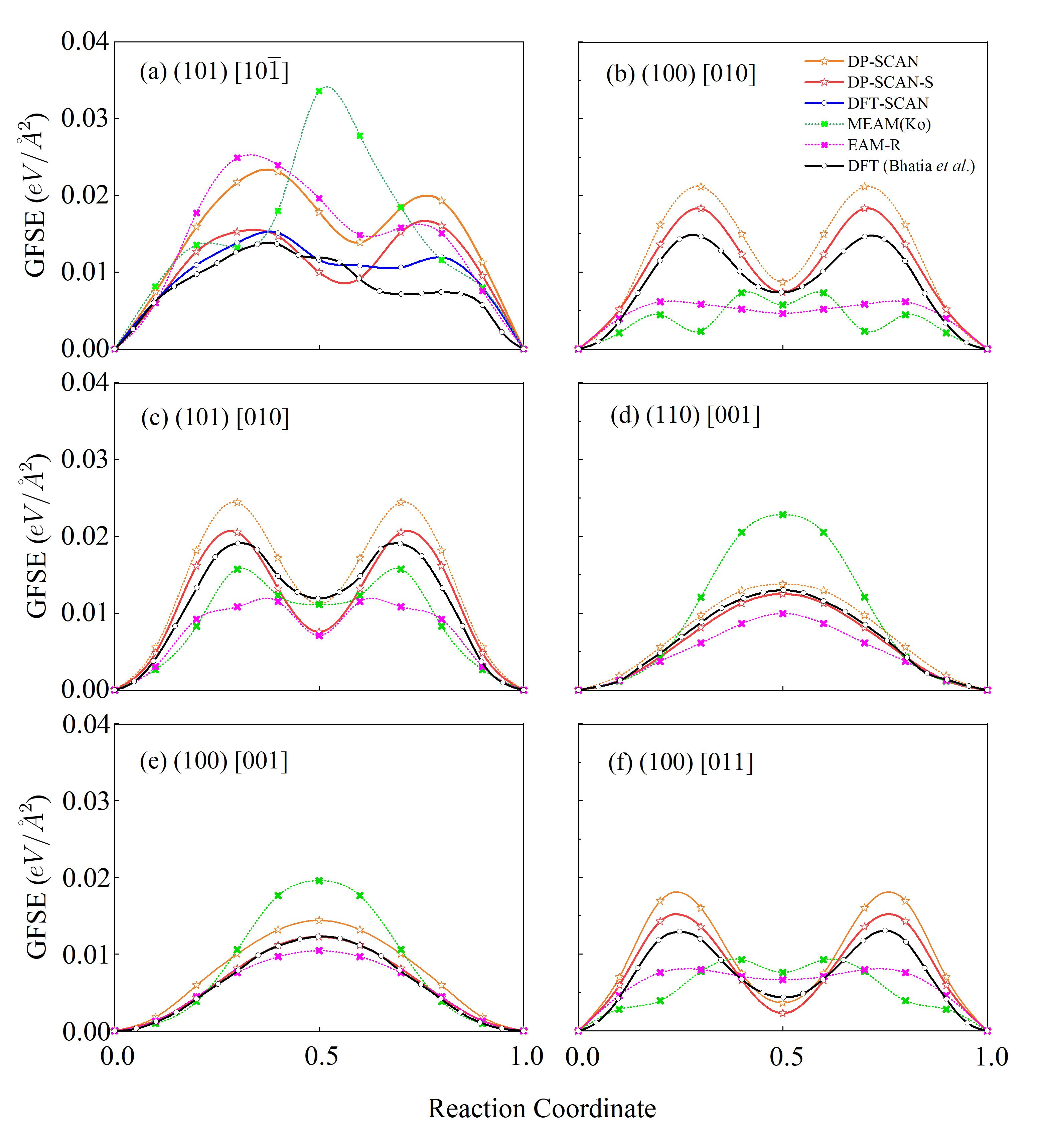}
    \caption{$\gamma$-line of 6 different slip systems of $\beta$-Sn by various potentials.} 
    \label{fig:SFE}
\end{figure}

We compute GFSEs for 6 slip systems of $\beta$-Sn, the configurations of which are shown in FIG.~\ref{fig:SYS}. The GFSE profiles ($\gamma$-line) are drawn in FIG.~\ref{fig:SFE}. We compare our results with those from MEAM (Ko), EAM-R and DFT calculations by Bhatia \textit{et al.}~\cite{ref15gammaBhatia}. The USFEs calculated by various  models are listed in TABLE~\ref{table:USFE}.

As shown in Fig.~\ref{fig:SFE}(a), none of the potential models were able to reproduce the $\gamma$-line of (101)[$10\overline{1}$] slip system, which exhibits a unique asymmetry validated by DFT with both PBE (Bhatia \textit{et al.}) and SCAN functionals (this work). 
As demonstrated in Figs.~\ref{fig:SFE}(b) to (f), DP-SCAN-S exhibits the highest overall accuracy in predicting $\gamma$-lines along various slip systems when compared to DFT calculations. 
This superior performance is further corroborated by the USFE values presented in TABLE~\ref{table:USFE}.
Note that the $\gamma$-line structure are not explicitly included in the training dataset,
which strongly highlight both the high generalization ability of the DP methods and the  importance of sufficient sampling of configurational space by the concurrent-learning scheme.

\renewcommand{\cellalign}{tl}
\begin{table}
\footnotesize
\centering
\caption{Unstable fault energies (USFE, in $\mathrm{meV/\mathrm{\AA}^2}$) obtained from the peak of curves in Fig.~\ref{fig:SFE}. Comparisons are made among results by MEAM (Ko)~\cite{ref11meam2018Ko}, ML potential EAM-R~\cite{ref12EAM-R}, DP-SCAN~\cite{ref1DP-SCAN}, DP-SCAN-S, DFT calculations by Bhatia \textit{et al.}~\cite{ref15gammaBhatia}, and DFT-SCAN (for (101)[10$\overline{1}$]). }
\label{table:USFE}
\setlength{\tabcolsep}{8pt}
\begin{tabular}{lllllll}
\hline
\hline
Slip Systems & DFT & DP-SCAN-S & DP-SCAN & EAM-R & MEAM (Ko) \\
\hline
(101)[10$\overline{1}$] & \makecell{15.1~(this work), \\13.7~\cite{ref15gammaBhatia}} & 15.3 & 23.1 & 24.9 & 33.6 \\
(100)[010] & 14.6~\cite{ref15gammaBhatia} & 18.3 & 21.1 & 6.1 & 7.3 \\
(101)[010] & 19.1~\cite{ref15gammaBhatia} & 20.5 & 24.4 & 11.5 & 15.6 \\
(110)[001] & 13.0~\cite{ref15gammaBhatia} & 12.5 & 13.8 & 10.0 & 22.8 \\
(100)[001] & 12.3~\cite{ref15gammaBhatia} & 12.3 & 14.4 & 10.4 & 19.6 \\
(100)[011] & 13.1~\cite{ref15gammaBhatia} & 15.2 & 18.1 & 8.0 & 9.3 \\
\hline
\end{tabular}
\end{table}

\subsection{Shock Hugoniot}

Hugoniot curves~\cite{Needham2018} describe the equation of state under shock wave. 
The Hugoniot curves of Sn have been calculated in previous works to estimate the ability of potentials to describe shock loaded Sn~\cite{ref6DandS-microjetting,TANG2020103479}.
We adopt the multi-scale shock technique (MSST)~\cite{ref14MSST} method implemented in the LAMMPS package and perform simulations to calculate Hugoniot using DP-SCAN-S. 
%
%
%
%
The initial system is $\beta$-Sn, which contains 13,500 atoms and scales 88.5 $\mathrm{\AA}$~in $x$ and $y$ direction and 48.14 $\mathrm{\AA}$~in $z$ direction. Then we relax the system for 10 ps under 300 K and 0 pressure in the NPT ensemble. After relaxation, shock wave propagates along the $z$ direction and the [001] crystal orientation. The MSST simulations are also performed using EAM-R. 

\begin{figure}
    \centering
    \includegraphics[width=.8\textwidth]{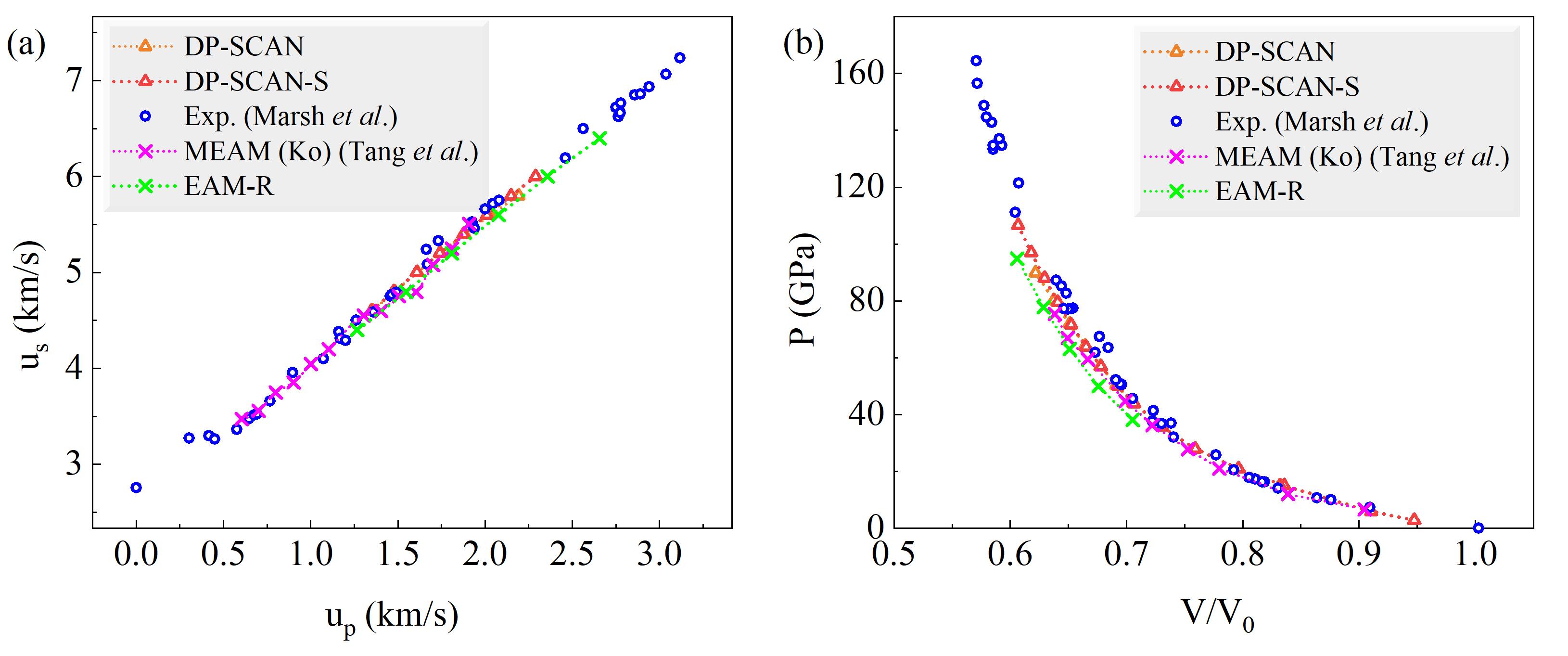}
    \caption{(a) Hugoniot curves on the $V$-$P$ plane, where $V$ is represented by the compress ratio $V/V_{0}$ and $P$ is the pressure in GPa. (b) Hugoniot curves of Sn on the u$_s$-u$_p$ plane, where u$_s$ and u$_p$ represent the shock wave velocity and particle velocity, respectively. Results from the DP-SCAN-S, DP-SCAN, MEAM (Ko)~\cite{TANG2020103479}, and EAM-R models are in red, orange, purple, and green, respectively.
    The experimental results~\cite{ref13HuExp} are in blue.}
    \label{fig:Hugoniot}
\end{figure} 

The Hugoniot curve, represented by the u$_p$-u$_s$ relationship and the V/V$_{0}$-P relationship, are depicted in Fig.~\ref{fig:Hugoniot} (a) and (b), respectively.
Here u$_p$ and u$_s$ represent the particle velocity and the shock wave velocity, respectively.
Hugoniot curve described by MEAM (Ko) are performed by Tang \textit{et al.}~\cite{TANG2020103479}. 
Fig.~\ref{fig:Hugoniot} (a) illustrates that the u$_p$-u$_s$ curves by various models exhibit a consistent linear relationship, which agrees well with both theoretical analysis~\cite{Needham2018} and experiment data~\cite{ref13HuExp}. 
Fig.~\ref{fig:Hugoniot} (b) shows that the Hugoniot pressure calculated by the DP-SCAN-S model exhibits the highest degree of proximity to the experimental results. Notably, the outcomes from other models align well with the experimental data below 50 GPa; however, as the pressure surpasses 50 GPa, more pronounced deviations emerge. Importantly, among all the models under investigation, the DP-SCAN-S model demonstrates the best agreement with the experimental findings, primarily because its data covers the broadest range of thermodynamic conditions.

\subsection{ Phase diagram and melting points}

\begin{figure}
    \includegraphics[width=.5\textwidth]{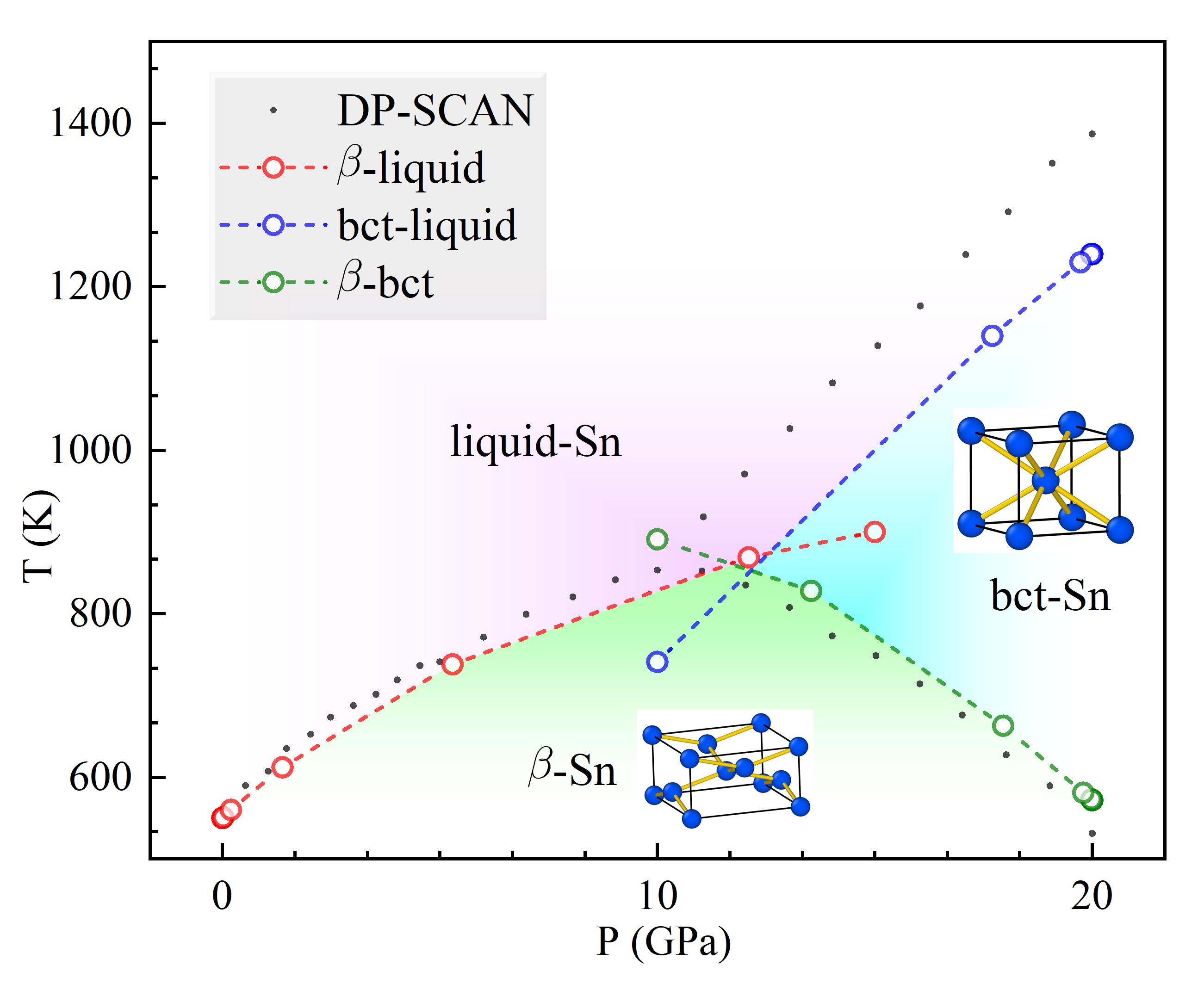}
    \caption{Phase diagram of Sn around $\beta$-bct-liquid triple point calculated by TI and GDI. Red circles and lines represent the $\beta$-liquid boundary. Green circles and lines represent the $\beta$-bct boundary. Blue circles and lines represent the bct-liquid boundary. Black dots represent the boundaries calculated with DP-SCAN. The triple point is predicted at (11.5 GPa, 860 K) by DP-SCAN-S, which is slightly right shifted compared to that of DP-SCAN at (10 GPa, 853.5 K).}
    \label{fig:TI}
\end{figure}

The accuracy of DP-SCAN-S in describing the phases of Sn is evaluated across three aspects: the pressure-temperature (P-T) phase diagram near the $\beta$-bct-liquid triple point, the enthalpy curve within a pressure range of 0 to 50 GPa, and the melting curve up to 90 GPa. Phase diagram around the $\beta$-bct-liquid triple point is shown in Fig.~\ref{fig:TI}. 
The relative Gibbs free energy difference with the reference system at a given pressure-temperature condition, ($P_0,T_0$), is firstly calculated via the Hamiltonian thermodynamic integration (HTI). 
At $P_0=0$ GPa, $T_0$ for the $\beta$ phase and liquid phase are 400~K and 700~K. 
At $P_0=20$ GPa, $T_0$ for the $\beta$ phase, $bct$ phase and liquid phase are 500~K, 900~K and 1300~K. 
Subsequently, the thermodynamic integration (TI) is used to determine the temperature dependence of the Gibbs free energy.
Phase separation points at specified pressures are identified by the intersection of the free energy versus temperature curves for the two involved phases.
Then the phase boundary are outlined using the Gibbs-Duhem integration (GDI) from the phase separation points.
The $\beta$-liquid phase boundary is integrated from 0 GPa to 15 GPa, and the $bct$-liquid phase boundary and $\beta$-$bct$ phase boundary is integrated from 20 GPa to 10 GPa.
The HTI, TI and GDI are calculated using the DPTI code, whose principles can be found in Ref.\cite{ref1DP-SCAN}.
The intersection point of the three boundaries corresponds to the $\beta$-bct-liquid triple point. 
DP-SCAN-S predicts the triple point at around (11.5 GPa, 860 K), which is close to the result of DP-SCAN at (10 GPa, 853.5 K) but slightly shifted towards higher pressure, though there is still a  deviation from the experimental triple point at (3.02 GPa, 562 K)~\cite{ref17MPXu}.

\begin{figure}
    \includegraphics[width=.5\textwidth]{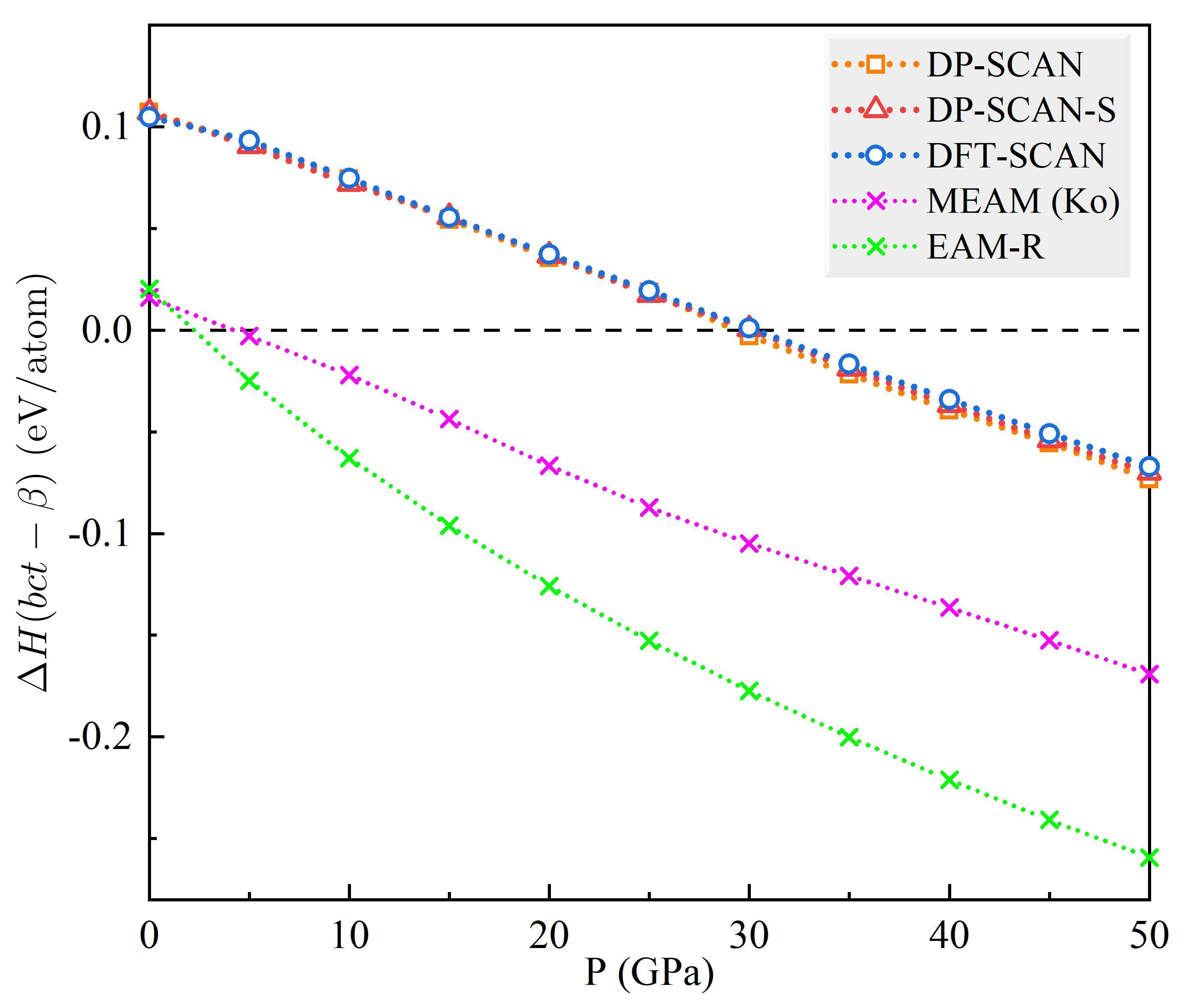}
    \caption{Enthalpy differences $\Delta H$ between the bct-Sn and $\beta$-Sn phases calculated at 0 K and 0-50 GPa. DP-SCAN-S and DP-SCAN agrees well with DFT-SCAN, while MEAM and EAM-R give a significantly underestimated phase transition point at 0~K.}
    \label{fig:EN}
\end{figure}

Enthalpy is a crucial state function used to describe the thermodynamic properties of a system. It is defined as $H=U+PV$ with $U$ being the internal energy of the system while $P$ and $V$ depict the pressure and volume of the system, respectively. At zero temperature, enthalpy is equal to Helmholtz free energy and can serve as a criterion for phase stability. 
Here, we adopt five models including DFT-SCAN, DP-SCAN, DP-SCAN-S, MEAM, and EAM-R to calculate the enthalpy difference $\Delta H$ between the $\beta$ and $bct$ phases and the results are shown in Fig.~\ref{fig:EN}.
To obtain the enthalpy, the conjugated gradient method is used to optimize the $\beta$ and bct structures at pressures below 50 GPa. 
We find the DP-SCAN-S, DP-SCAN and DFT-SCAN models yield close results. In addition, the phase boundary of around 30 GPa at 0 K is also consistent with the trend extrapolated from experimental results in Fig.~\ref{fig:TI}. 
The transition point at 0~K calculated by the MEAM (Ko) and EAM-R models is lower than the DFT-SCAN results. 
The EAM-R and MEAM(Ko) methods predict that the bct phase is more stable than the $\beta$ phase at above 3 and 6 GPa, respectively.
These predictions are significantly lower than the transition point extrapolated from the experimental results\cite{ref1DP-SCAN}.

\begin{figure}
    \includegraphics[width=.5\textwidth]{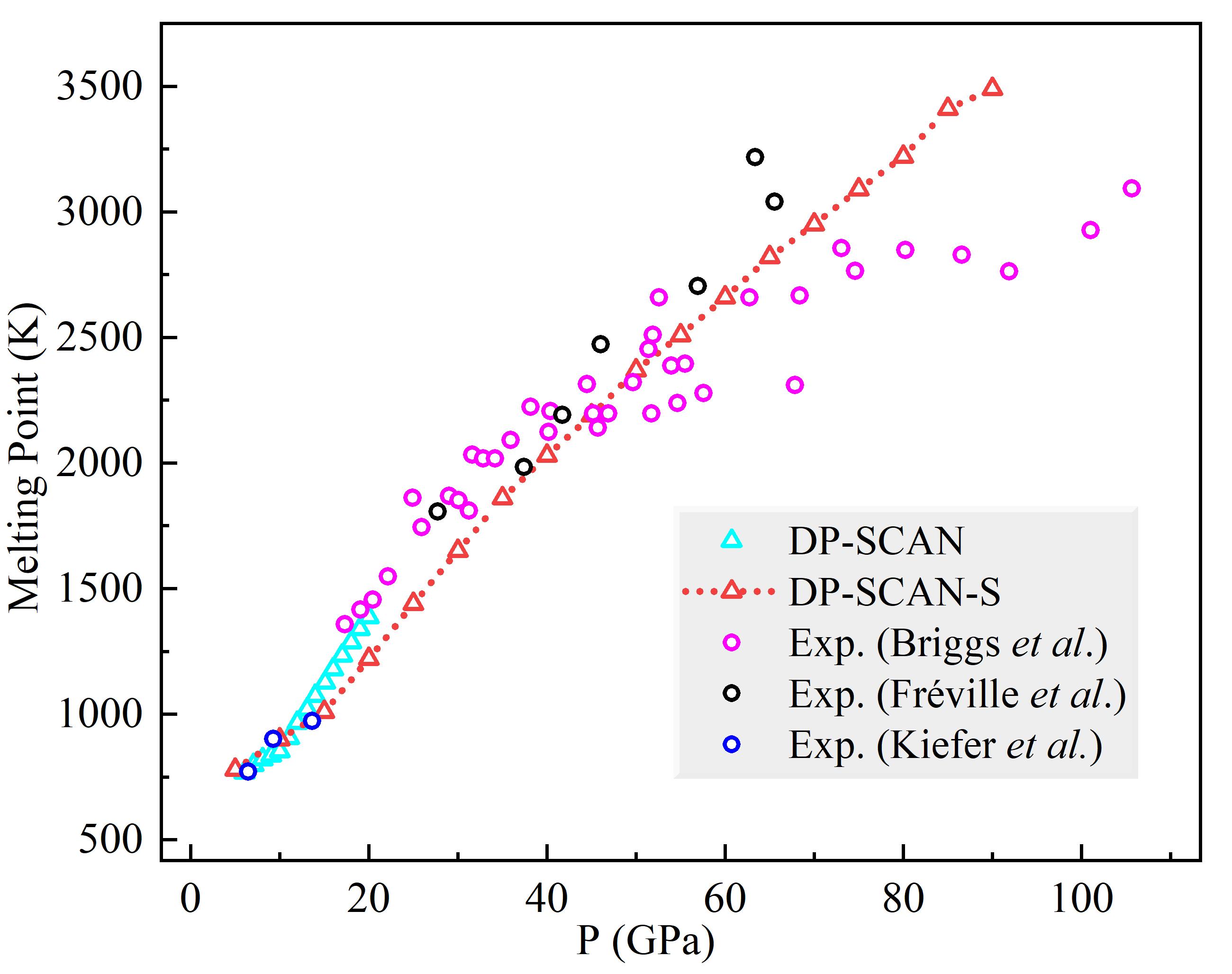}
    \caption{Predicted melting points of Sn at high pressures and available experimental data from Kiefer et al.~\cite{kiefer2002melting}, Briggs \textit{et al.}~\cite{ref16MPBriggs}, and Fr\'eville \textit{et al.}~\cite{PhysRevB.109.104116}.
  For the predictions of DP-SCAN-S, different initial states are used under different pressures, $\beta$-Sn under 5-10 GPa, bct under 15-35 GPa, and bcc under 40 GPa above. }
    \label{fig:MP}
\end{figure}

%
We perform solid-liquid co-existence simulations with DP-SCAN-S to determine the melting points of Sn under various pressures.
Each configuration contains 4096 atoms, evenly divided between solid and liquid phases. 
The configurations were constructed by the following steps. First, the solid phase containing 2048 atoms was relaxed at pressures ranging from 0 to 90 GPa. Next, the solid phase was heated under the NPT ensemble for 10 ps to mimic the melting process and obtain the liquid structure, with the box size fixed in the $x$ and $y$ directions. Finally, the solid and liquid structures were combined in the $z$ direction.
The solid-liquid coexistence systems were equilibrated under the NPT ensemble for 50 ps with a timestep of 1.0 fs. The pressure increment between successive simulations is 5 GPa, while the temperature interval in each of the different simulations is 10 K.
Fig.~\ref{fig:MP} shows that the melting points obtained by the DP-SCAN-S model generally reproduce the experimental results at both low and high pressures.
Below 10 GPa, the $\beta$-Sn structure is adopted as the initial state. 
From 15 to 35 GPa, the bct-Sn structure is utilized, while above 35 GPa, the bcc structures are employed. 
The transition of the initial state from $\beta$-Sn to bct-Sn leads to a change in the slope at 15 GPa. 
At lower pressures, the melting points predicted by DP-SCAN-S is in excellent agreement with the experimental results by Kiefer et al.\cite{kiefer2002melting}
At pressure between 20 and 60 GPa, the melting point predicted by DP-SCAN-S shows an overall good agreement with the experimental studies \cite{ref16MPBriggs,PhysRevB.109.104116}.
Above 60 GPa, however, the experimental studies by Briggs et al.~\cite{ref16MPBriggs} shows a saturation of melting point with pressure, though DP-SCAN-S does not reproduce such tendency.
Notably, a more advanced study recently by Fr\'eville {et al.}~\cite{PhysRevB.109.104116} shows no change of slope at higher pressures.

The good agreement of melting point ensures DP-SCAN-S model to accurately capture the dynamic melting behaviors during loading and unloading of compresive shock-waves.


\section{Conclusions}

In conclusion, we developed a machine learning potential for Sn specifically designed for use in dynamic simulations of shock responses.
The data were collected through DP-GEN workflow over a wide range of thermodynamic conditions, spanning pressures from 0 to 100 GPa and temperatures from 0 to 5000 K. 
Configurations with free surfaces were added into training data to enhance its ability to describe fracture behaviors. 
The collected data were labeled by DFT calculations with SCAN exchange-correlation functional. 

The model, DP-SCAN-S achieves the accuracy of DFT in various properties including cohesive energy, lattice constants, elastic constants, low index surface energy, and vacancy formation energy. Moreover, it reproduces the relative enthalpy for $\beta$ and bct phases below 50 GPa.
DP-SCAN-S achieves the best agreement to DFT results on the prediction of $\gamma$-lines among all tested potential.
At finite temperature, DP-SCAN-S accurately predict the P-T phase diagram of Sn near the triple point, as well as high-pressure melting curves. 
Among the machine learning potential models examined, the Hugoniot curve in the pressure-volume (P-V) plane calculated by DP-SCAN-S showed the closest agreement with experimental data. 
Therefore, we anticipate that the DP-SCAN-S model serves as a reliable interatomic potential suitable for simulating the dynamic mechanical behavior of Sn under shock loading.


\nocite{*}
\bibliography{main} 

\providecommand{\noopsort}[1]{}\providecommand{\singleletter}[1]{#1}%
\begin{thebibliography}{97}%
\makeatletter
\providecommand \@ifxundefined [1]{%
 \@ifx{#1\undefined}
}%
\providecommand \@ifnum [1]{%
 \ifnum #1\expandafter \@firstoftwo
 \else \expandafter \@secondoftwo
 \fi
}%
\providecommand \@ifx [1]{%
 \ifx #1\expandafter \@firstoftwo
 \else \expandafter \@secondoftwo
 \fi
}%
\providecommand \natexlab [1]{#1}%
\providecommand \enquote  [1]{``#1''}%
\providecommand \bibnamefont  [1]{#1}%
\providecommand \bibfnamefont [1]{#1}%
\providecommand \citenamefont [1]{#1}%
\providecommand \href@noop [0]{\@secondoftwo}%
\providecommand \href [0]{\begingroup \@sanitize@url \@href}%
\providecommand \@href[1]{\@@startlink{#1}\@@href}%
\providecommand \@@href[1]{\endgroup#1\@@endlink}%
\providecommand \@sanitize@url [0]{\catcode `\\12\catcode `\$12\catcode `\&12\catcode `\#12\catcode `\^12\catcode `\_12\catcode `\%12\relax}%
\providecommand \@@startlink[1]{}%
\providecommand \@@endlink[0]{}%
\providecommand \url  [0]{\begingroup\@sanitize@url \@url }%
\providecommand \@url [1]{\endgroup\@href {#1}{\urlprefix }}%
\providecommand \urlprefix  [0]{URL }%
\providecommand \Eprint [0]{\href }%
\providecommand \doibase [0]{https://doi.org/}%
\providecommand \selectlanguage [0]{\@gobble}%
\providecommand \bibinfo  [0]{\@secondoftwo}%
\providecommand \bibfield  [0]{\@secondoftwo}%
\providecommand \translation [1]{[#1]}%
\providecommand \BibitemOpen [0]{}%
\providecommand \bibitemStop [0]{}%
\providecommand \bibitemNoStop [0]{.\EOS\space}%
\providecommand \EOS [0]{\spacefactor3000\relax}%
\providecommand \BibitemShut  [1]{\csname bibitem#1\endcsname}%
\let\auto@bib@innerbib\@empty
\bibitem [{\citenamefont {Novikov}(1981)}]{Novikov1981}%
  \BibitemOpen
  \bibfield  {author} {\bibinfo {author} {\bibfnamefont {S.~A.}\ \bibnamefont {Novikov}},\ }\bibfield  {title} {\bibinfo {title} {Shear stress and spall strength of materials under shock loads (review)},\ }\href {https://doi.org/https://doi.org/10.1007/BF00907567} {\bibfield  {journal} {\bibinfo  {journal} {Journal of Applied Mechanics and Technical Physics}\ }\textbf {\bibinfo {volume} {22}},\ \bibinfo {pages} {385} (\bibinfo {year} {1981})}\BibitemShut {NoStop}%
\bibitem [{\citenamefont {Duvall}\ and\ \citenamefont {Graham}(1977)}]{RevModPhys.49.523}%
  \BibitemOpen
  \bibfield  {author} {\bibinfo {author} {\bibfnamefont {G.~E.}\ \bibnamefont {Duvall}}\ and\ \bibinfo {author} {\bibfnamefont {R.~A.}\ \bibnamefont {Graham}},\ }\bibfield  {title} {\bibinfo {title} {Phase transitions under shock-wave loading},\ }\href {https://doi.org/10.1103/RevModPhys.49.523} {\bibfield  {journal} {\bibinfo  {journal} {Rev. Mod. Phys.}\ }\textbf {\bibinfo {volume} {49}},\ \bibinfo {pages} {523} (\bibinfo {year} {1977})}\BibitemShut {NoStop}%
\bibitem [{\citenamefont {Bringa}\ \emph {et~al.}(2006)\citenamefont {Bringa}, \citenamefont {Rosolankova}, \citenamefont {Rudd}, \citenamefont {Remington}, \citenamefont {Wark}, \citenamefont {Duchaineau}, \citenamefont {Kalantar}, \citenamefont {Hawreliak},\ and\ \citenamefont {Belak}}]{NatureMaterials2006}%
  \BibitemOpen
  \bibfield  {author} {\bibinfo {author} {\bibfnamefont {E.~M.}\ \bibnamefont {Bringa}}, \bibinfo {author} {\bibfnamefont {K.}~\bibnamefont {Rosolankova}}, \bibinfo {author} {\bibfnamefont {R.~E.}\ \bibnamefont {Rudd}}, \bibinfo {author} {\bibfnamefont {B.~A.}\ \bibnamefont {Remington}}, \bibinfo {author} {\bibfnamefont {J.~S.}\ \bibnamefont {Wark}}, \bibinfo {author} {\bibfnamefont {M.}~\bibnamefont {Duchaineau}}, \bibinfo {author} {\bibfnamefont {D.~H.}\ \bibnamefont {Kalantar}}, \bibinfo {author} {\bibfnamefont {J.}~\bibnamefont {Hawreliak}},\ and\ \bibinfo {author} {\bibfnamefont {J.}~\bibnamefont {Belak}},\ }\bibfield  {title} {\bibinfo {title} {Shock deformation of face-centred-cubic metals on subnanosecond timescales},\ }\href {https://doi.org/10.1038/nmat1735} {\bibfield  {journal} {\bibinfo  {journal} {Nature Materials}\ }\textbf {\bibinfo {volume} {5}},\ \bibinfo {pages} {805} (\bibinfo {year} {2006})}\BibitemShut {NoStop}%
\bibitem [{\citenamefont {De~Ress{\'e}guier}\ \emph {et~al.}(2007{\natexlab{a}})\citenamefont {De~Ress{\'e}guier}, \citenamefont {Signor}, \citenamefont {Dragon}, \citenamefont {Boustie}, \citenamefont {Roy},\ and\ \citenamefont {Llorca}}]{de2007experimental}%
  \BibitemOpen
  \bibfield  {author} {\bibinfo {author} {\bibfnamefont {T.}~\bibnamefont {De~Ress{\'e}guier}}, \bibinfo {author} {\bibfnamefont {L.}~\bibnamefont {Signor}}, \bibinfo {author} {\bibfnamefont {A.}~\bibnamefont {Dragon}}, \bibinfo {author} {\bibfnamefont {M.}~\bibnamefont {Boustie}}, \bibinfo {author} {\bibfnamefont {G.}~\bibnamefont {Roy}},\ and\ \bibinfo {author} {\bibfnamefont {F.}~\bibnamefont {Llorca}},\ }\bibfield  {title} {\bibinfo {title} {Experimental investigation of liquid spall in laser shock-loaded tin},\ }\href@noop {} {\bibfield  {journal} {\bibinfo  {journal} {Journal of applied physics}\ }\textbf {\bibinfo {volume} {101}} (\bibinfo {year} {2007}{\natexlab{a}})}\BibitemShut {NoStop}%
\bibitem [{\citenamefont {De~Ress{\'e}guier}\ \emph {et~al.}(2007{\natexlab{b}})\citenamefont {De~Ress{\'e}guier}, \citenamefont {Signor}, \citenamefont {Dragon}, \citenamefont {Severin},\ and\ \citenamefont {Boustie}}]{de2007spallation}%
  \BibitemOpen
  \bibfield  {author} {\bibinfo {author} {\bibfnamefont {T.}~\bibnamefont {De~Ress{\'e}guier}}, \bibinfo {author} {\bibfnamefont {L.}~\bibnamefont {Signor}}, \bibinfo {author} {\bibfnamefont {A.}~\bibnamefont {Dragon}}, \bibinfo {author} {\bibfnamefont {P.}~\bibnamefont {Severin}},\ and\ \bibinfo {author} {\bibfnamefont {M.}~\bibnamefont {Boustie}},\ }\bibfield  {title} {\bibinfo {title} {Spallation in laser shock-loaded tin below and just above melting on release},\ }\href@noop {} {\bibfield  {journal} {\bibinfo  {journal} {Journal of Applied Physics}\ }\textbf {\bibinfo {volume} {102}} (\bibinfo {year} {2007}{\natexlab{b}})}\BibitemShut {NoStop}%
\bibitem [{\citenamefont {Ashitkov}\ \emph {et~al.}(2016)\citenamefont {Ashitkov}, \citenamefont {Komarov}, \citenamefont {Ovchinnikov}, \citenamefont {Struleva},\ and\ \citenamefont {Agranat}}]{Ashitkov}%
  \BibitemOpen
  \bibfield  {author} {\bibinfo {author} {\bibfnamefont {S.~I.}\ \bibnamefont {Ashitkov}}, \bibinfo {author} {\bibfnamefont {P.~S.}\ \bibnamefont {Komarov}}, \bibinfo {author} {\bibfnamefont {A.~V.}\ \bibnamefont {Ovchinnikov}}, \bibinfo {author} {\bibfnamefont {E.~V.}\ \bibnamefont {Struleva}},\ and\ \bibinfo {author} {\bibfnamefont {M.~B.}\ \bibnamefont {Agranat}},\ }\bibfield  {title} {\bibinfo {title} {Strength of liquid tin at extremely high strain rates under a femtosecond laser action},\ }\href {https://doi.org/10.1134/S0021364016080038} {\bibfield  {journal} {\bibinfo  {journal} {JETP Letters}\ }\textbf {\bibinfo {volume} {103}},\ \bibinfo {pages} {544} (\bibinfo {year} {2016})}\BibitemShut {NoStop}%
\bibitem [{\citenamefont {Kanel}\ \emph {et~al.}(2015)\citenamefont {Kanel}, \citenamefont {Savinykh}, \citenamefont {Garkushin},\ and\ \citenamefont {Razorenov}}]{Kanel2015}%
  \BibitemOpen
  \bibfield  {author} {\bibinfo {author} {\bibfnamefont {G.~I.}\ \bibnamefont {Kanel}}, \bibinfo {author} {\bibfnamefont {A.~S.}\ \bibnamefont {Savinykh}}, \bibinfo {author} {\bibfnamefont {G.~V.}\ \bibnamefont {Garkushin}},\ and\ \bibinfo {author} {\bibfnamefont {S.~V.}\ \bibnamefont {Razorenov}},\ }\bibfield  {title} {\bibinfo {title} {Dynamic strength of tin and lead melts},\ }\href {https://doi.org/10.1134/S0021364015200059} {\bibfield  {journal} {\bibinfo  {journal} {JETP Letters}\ }\textbf {\bibinfo {volume} {102}},\ \bibinfo {pages} {548} (\bibinfo {year} {2015})}\BibitemShut {NoStop}%
\bibitem [{\citenamefont {Smith}(1985)}]{SMITH198569}%
  \BibitemOpen
  \bibfield  {author} {\bibinfo {author} {\bibfnamefont {R.~W.}\ \bibnamefont {Smith}},\ }\bibfield  {title} {\bibinfo {title} {The $\alpha$(semiconductor) ag $\beta$(metal) transition in tin},\ }\href {https://doi.org/https://doi.org/10.1016/0022-5088(85)90391-1} {\bibfield  {journal} {\bibinfo  {journal} {Journal of the Less Common Metals}\ }\textbf {\bibinfo {volume} {114}},\ \bibinfo {pages} {69} (\bibinfo {year} {1985})}\BibitemShut {NoStop}%
\bibitem [{\citenamefont {Liu}\ and\ \citenamefont {Liu}(1986)}]{Liu1986CompressionsAP}%
  \BibitemOpen
  \bibfield  {author} {\bibinfo {author} {\bibfnamefont {M.}~\bibnamefont {Liu}}\ and\ \bibinfo {author} {\bibfnamefont {L.-G.}\ \bibnamefont {Liu}},\ }\bibfield  {title} {\bibinfo {title} {Compressions and phase transitions of tin to half a megabar},\ }\href {https://api.semanticscholar.org/CorpusID:99665311} {\bibfield  {journal} {\bibinfo  {journal} {High Temperatures-high Pressures}\ }\textbf {\bibinfo {volume} {18}},\ \bibinfo {pages} {79} (\bibinfo {year} {1986})}\BibitemShut {NoStop}%
\bibitem [{\citenamefont {Olijnyk}\ and\ \citenamefont {Holzapfel}(1984)}]{olijnyk1984phase}%
  \BibitemOpen
  \bibfield  {author} {\bibinfo {author} {\bibfnamefont {H.}~\bibnamefont {Olijnyk}}\ and\ \bibinfo {author} {\bibfnamefont {W.}~\bibnamefont {Holzapfel}},\ }\bibfield  {title} {\bibinfo {title} {Phase transitions in si, ge and sn under pressure},\ }\href@noop {} {\bibfield  {journal} {\bibinfo  {journal} {Le Journal de Physique Colloques}\ }\textbf {\bibinfo {volume} {45}},\ \bibinfo {pages} {C8} (\bibinfo {year} {1984})}\BibitemShut {NoStop}%
\bibitem [{\citenamefont {Salamat}\ \emph {et~al.}(2011)\citenamefont {Salamat}, \citenamefont {Garbarino}, \citenamefont {Dewaele}, \citenamefont {Bouvier}, \citenamefont {Petitgirard}, \citenamefont {Pickard}, \citenamefont {McMillan},\ and\ \citenamefont {Mezouar}}]{salamat2011dense}%
  \BibitemOpen
  \bibfield  {author} {\bibinfo {author} {\bibfnamefont {A.}~\bibnamefont {Salamat}}, \bibinfo {author} {\bibfnamefont {G.}~\bibnamefont {Garbarino}}, \bibinfo {author} {\bibfnamefont {A.}~\bibnamefont {Dewaele}}, \bibinfo {author} {\bibfnamefont {P.}~\bibnamefont {Bouvier}}, \bibinfo {author} {\bibfnamefont {S.}~\bibnamefont {Petitgirard}}, \bibinfo {author} {\bibfnamefont {C.~J.}\ \bibnamefont {Pickard}}, \bibinfo {author} {\bibfnamefont {P.~F.}\ \bibnamefont {McMillan}},\ and\ \bibinfo {author} {\bibfnamefont {M.}~\bibnamefont {Mezouar}},\ }\bibfield  {title} {\bibinfo {title} {Dense close-packed phase of tin above 157 gpa observed experimentally via angle-dispersive x-ray diffraction},\ }\href@noop {} {\bibfield  {journal} {\bibinfo  {journal} {Physical Review B}\ }\textbf {\bibinfo {volume} {84}},\ \bibinfo {pages} {140104} (\bibinfo {year} {2011})}\BibitemShut {NoStop}%
\bibitem [{\citenamefont {Briggs}\ \emph {et~al.}(2019{\natexlab{a}})\citenamefont {Briggs}, \citenamefont {Torchio}, \citenamefont {Sollier}, \citenamefont {Occelli}, \citenamefont {Videau}, \citenamefont {Kretzschmar},\ and\ \citenamefont {Wulff}}]{Brigs2019Observation}%
  \BibitemOpen
  \bibfield  {author} {\bibinfo {author} {\bibfnamefont {R.}~\bibnamefont {Briggs}}, \bibinfo {author} {\bibfnamefont {R.}~\bibnamefont {Torchio}}, \bibinfo {author} {\bibfnamefont {A.}~\bibnamefont {Sollier}}, \bibinfo {author} {\bibfnamefont {F.}~\bibnamefont {Occelli}}, \bibinfo {author} {\bibfnamefont {L.}~\bibnamefont {Videau}}, \bibinfo {author} {\bibfnamefont {N.}~\bibnamefont {Kretzschmar}},\ and\ \bibinfo {author} {\bibfnamefont {M.}~\bibnamefont {Wulff}},\ }\bibfield  {title} {\bibinfo {title} {Observation of the shock-induced $\beta$-sn to b.c.t.-sn transition using time-resolved x-ray diffraction},\ }\href {https://doi.org/10.1107/s1600577518015059} {\bibfield  {journal} {\bibinfo  {journal} {Journal of Synchrotron Radiation}\ }\textbf {\bibinfo {volume} {26}},\ \bibinfo {pages} {96} (\bibinfo {year} {2019}{\natexlab{a}})}\BibitemShut {NoStop}%
\bibitem [{\citenamefont {Hu}\ \emph {et~al.}(2008)\citenamefont {Hu}, \citenamefont {Zhou}, \citenamefont {Dai}, \citenamefont {Tan},\ and\ \citenamefont {Li}}]{hu2008shock}%
  \BibitemOpen
  \bibfield  {author} {\bibinfo {author} {\bibfnamefont {J.}~\bibnamefont {Hu}}, \bibinfo {author} {\bibfnamefont {X.}~\bibnamefont {Zhou}}, \bibinfo {author} {\bibfnamefont {C.}~\bibnamefont {Dai}}, \bibinfo {author} {\bibfnamefont {H.}~\bibnamefont {Tan}},\ and\ \bibinfo {author} {\bibfnamefont {J.}~\bibnamefont {Li}},\ }\bibfield  {title} {\bibinfo {title} {Shock-induced bct-bcc transition and melting of tin identified by sound velocity measurements},\ }\href@noop {} {\bibfield  {journal} {\bibinfo  {journal} {Journal of Applied Physics}\ }\textbf {\bibinfo {volume} {104}} (\bibinfo {year} {2008})}\BibitemShut {NoStop}%
\bibitem [{\citenamefont {Signor}\ \emph {et~al.}(2009)\citenamefont {Signor}, \citenamefont {Roy}, \citenamefont {Chanal}, \citenamefont {Héreil}, \citenamefont {Buy}, \citenamefont {Voltz}, \citenamefont {Llorca}, \citenamefont {De~Rességuier},\ and\ \citenamefont {Dragon}}]{Signor2009}%
  \BibitemOpen
  \bibfield  {author} {\bibinfo {author} {\bibfnamefont {L.}~\bibnamefont {Signor}}, \bibinfo {author} {\bibfnamefont {G.}~\bibnamefont {Roy}}, \bibinfo {author} {\bibfnamefont {P.}~\bibnamefont {Chanal}}, \bibinfo {author} {\bibfnamefont {P.}~\bibnamefont {Héreil}}, \bibinfo {author} {\bibfnamefont {F.}~\bibnamefont {Buy}}, \bibinfo {author} {\bibfnamefont {C.}~\bibnamefont {Voltz}}, \bibinfo {author} {\bibfnamefont {F.}~\bibnamefont {Llorca}}, \bibinfo {author} {\bibfnamefont {T.}~\bibnamefont {De~Rességuier}},\ and\ \bibinfo {author} {\bibfnamefont {A.}~\bibnamefont {Dragon}},\ }\bibfield  {title} {\bibinfo {title} {Debris cloud ejection from shock‐loaded tin melted on release or on compression},\ }\href {https://doi.org/10.1063/1.3294984} {\bibfield  {journal} {\bibinfo  {journal} {AIP Conference Proceedings}\ }\textbf {\bibinfo {volume} {1195}},\ \bibinfo {pages} {1065} (\bibinfo {year} {2009})}\BibitemShut {NoStop}%
\bibitem [{\citenamefont {Kadau}\ \emph {et~al.}(2002{\natexlab{a}})\citenamefont {Kadau}, \citenamefont {Germann}, \citenamefont {Lomdahl},\ and\ \citenamefont {Holian}}]{kadau2002microscopic}%
  \BibitemOpen
  \bibfield  {author} {\bibinfo {author} {\bibfnamefont {K.}~\bibnamefont {Kadau}}, \bibinfo {author} {\bibfnamefont {T.~C.}\ \bibnamefont {Germann}}, \bibinfo {author} {\bibfnamefont {P.~S.}\ \bibnamefont {Lomdahl}},\ and\ \bibinfo {author} {\bibfnamefont {B.~L.}\ \bibnamefont {Holian}},\ }\bibfield  {title} {\bibinfo {title} {Microscopic view of structural phase transitions induced by shock waves},\ }\href@noop {} {\bibfield  {journal} {\bibinfo  {journal} {science}\ }\textbf {\bibinfo {volume} {296}},\ \bibinfo {pages} {1681} (\bibinfo {year} {2002}{\natexlab{a}})}\BibitemShut {NoStop}%
\bibitem [{\citenamefont {Hwang}\ \emph {et~al.}(2020)\citenamefont {Hwang}, \citenamefont {Galtier}, \citenamefont {Cynn}, \citenamefont {Eom}, \citenamefont {Chun}, \citenamefont {Bang}, \citenamefont {Hwang}, \citenamefont {Choi}, \citenamefont {Kim}, \citenamefont {Kong}, \citenamefont {Kwon}, \citenamefont {Kang}, \citenamefont {Lee}, \citenamefont {Park}, \citenamefont {Lee}, \citenamefont {Lee}, \citenamefont {Yang}, \citenamefont {Shim}, \citenamefont {Vogt}, \citenamefont {Kim}, \citenamefont {Park}, \citenamefont {Kim}, \citenamefont {Nam}, \citenamefont {Lee}, \citenamefont {Hyun}, \citenamefont {Kim}, \citenamefont {Koo}, \citenamefont {Kao}, \citenamefont {Sekine},\ and\ \citenamefont {Lee}}]{doi:10.1126/sciadv.aaz5132}%
  \BibitemOpen
  \bibfield  {author} {\bibinfo {author} {\bibfnamefont {H.}~\bibnamefont {Hwang}}, \bibinfo {author} {\bibfnamefont {E.}~\bibnamefont {Galtier}}, \bibinfo {author} {\bibfnamefont {H.}~\bibnamefont {Cynn}}, \bibinfo {author} {\bibfnamefont {I.}~\bibnamefont {Eom}}, \bibinfo {author} {\bibfnamefont {S.~H.}\ \bibnamefont {Chun}}, \bibinfo {author} {\bibfnamefont {Y.}~\bibnamefont {Bang}}, \bibinfo {author} {\bibfnamefont {G.~C.}\ \bibnamefont {Hwang}}, \bibinfo {author} {\bibfnamefont {J.}~\bibnamefont {Choi}}, \bibinfo {author} {\bibfnamefont {T.}~\bibnamefont {Kim}}, \bibinfo {author} {\bibfnamefont {M.}~\bibnamefont {Kong}}, \bibinfo {author} {\bibfnamefont {S.}~\bibnamefont {Kwon}}, \bibinfo {author} {\bibfnamefont {K.}~\bibnamefont {Kang}}, \bibinfo {author} {\bibfnamefont {H.~J.}\ \bibnamefont {Lee}}, \bibinfo {author} {\bibfnamefont {C.}~\bibnamefont {Park}}, \bibinfo {author} {\bibfnamefont {J.~I.}\ \bibnamefont {Lee}}, \bibinfo {author} {\bibfnamefont {Y.}~\bibnamefont {Lee}}, \bibinfo {author}
  {\bibfnamefont {W.}~\bibnamefont {Yang}}, \bibinfo {author} {\bibfnamefont {S.-H.}\ \bibnamefont {Shim}}, \bibinfo {author} {\bibfnamefont {T.}~\bibnamefont {Vogt}}, \bibinfo {author} {\bibfnamefont {S.}~\bibnamefont {Kim}}, \bibinfo {author} {\bibfnamefont {J.}~\bibnamefont {Park}}, \bibinfo {author} {\bibfnamefont {S.}~\bibnamefont {Kim}}, \bibinfo {author} {\bibfnamefont {D.}~\bibnamefont {Nam}}, \bibinfo {author} {\bibfnamefont {J.~H.}\ \bibnamefont {Lee}}, \bibinfo {author} {\bibfnamefont {H.}~\bibnamefont {Hyun}}, \bibinfo {author} {\bibfnamefont {M.}~\bibnamefont {Kim}}, \bibinfo {author} {\bibfnamefont {T.-Y.}\ \bibnamefont {Koo}}, \bibinfo {author} {\bibfnamefont {C.-C.}\ \bibnamefont {Kao}}, \bibinfo {author} {\bibfnamefont {T.}~\bibnamefont {Sekine}},\ and\ \bibinfo {author} {\bibfnamefont {Y.}~\bibnamefont {Lee}},\ }\bibfield  {title} {\bibinfo {title} {Subnanosecond phase transition dynamics in laser-shocked iron},\ }\href {https://doi.org/10.1126/sciadv.aaz5132} {\bibfield  {journal} {\bibinfo
   {journal} {Science Advances}\ }\textbf {\bibinfo {volume} {6}},\ \bibinfo {pages} {eaaz5132} (\bibinfo {year} {2020})},\ \Eprint {https://arxiv.org/abs/https://www.science.org/doi/pdf/10.1126/sciadv.aaz5132} {https://www.science.org/doi/pdf/10.1126/sciadv.aaz5132} \BibitemShut {NoStop}%
\bibitem [{\citenamefont {Pandolfi}\ \emph {et~al.}(2022)\citenamefont {Pandolfi}, \citenamefont {Brown}, \citenamefont {Stubley}, \citenamefont {Higginbotham}, \citenamefont {Bolme}, \citenamefont {Lee}, \citenamefont {Nagler}, \citenamefont {Galtier}, \citenamefont {Sandberg}, \citenamefont {Yang}, \citenamefont {Mao}, \citenamefont {Wark},\ and\ \citenamefont {Gleason}}]{silvia2022NC}%
  \BibitemOpen
  \bibfield  {author} {\bibinfo {author} {\bibfnamefont {S.}~\bibnamefont {Pandolfi}}, \bibinfo {author} {\bibfnamefont {S.~B.}\ \bibnamefont {Brown}}, \bibinfo {author} {\bibfnamefont {P.~G.}\ \bibnamefont {Stubley}}, \bibinfo {author} {\bibfnamefont {A.}~\bibnamefont {Higginbotham}}, \bibinfo {author} {\bibfnamefont {C.~A.}\ \bibnamefont {Bolme}}, \bibinfo {author} {\bibfnamefont {H.~J.}\ \bibnamefont {Lee}}, \bibinfo {author} {\bibfnamefont {B.}~\bibnamefont {Nagler}}, \bibinfo {author} {\bibfnamefont {E.}~\bibnamefont {Galtier}}, \bibinfo {author} {\bibfnamefont {R.~L.}\ \bibnamefont {Sandberg}}, \bibinfo {author} {\bibfnamefont {W.}~\bibnamefont {Yang}}, \bibinfo {author} {\bibfnamefont {W.~L.}\ \bibnamefont {Mao}}, \bibinfo {author} {\bibfnamefont {J.~S.}\ \bibnamefont {Wark}},\ and\ \bibinfo {author} {\bibfnamefont {A.~E.}\ \bibnamefont {Gleason}},\ }\bibfield  {title} {\bibinfo {title} {Atomistic deformation mechanism of silicon under laser-driven shock compression},\ }\href
  {https://doi.org/10.1038/s41467-022-33220-0} {\bibfield  {journal} {\bibinfo  {journal} {Nature Communications}\ }\textbf {\bibinfo {volume} {13}},\ \bibinfo {pages} {5535} (\bibinfo {year} {2022})}\BibitemShut {NoStop}%
\bibitem [{\citenamefont {Holian}\ and\ \citenamefont {Lomdahl}(1998)}]{science1998}%
  \BibitemOpen
  \bibfield  {author} {\bibinfo {author} {\bibfnamefont {B.}~\bibnamefont {Holian}}\ and\ \bibinfo {author} {\bibfnamefont {P.}~\bibnamefont {Lomdahl}},\ }\bibfield  {title} {\bibinfo {title} {Plasticity induced by shock waves in nonequilibrium molecular-dynamics simulations},\ }\href {https://doi.org/10.1126/science.280.5372.2085} {\bibfield  {journal} {\bibinfo  {journal} {Science}\ }\textbf {\bibinfo {volume} {280}},\ \bibinfo {pages} {2086} (\bibinfo {year} {1998})}\BibitemShut {NoStop}%
\bibitem [{\citenamefont {Kadau}\ \emph {et~al.}(2002{\natexlab{b}})\citenamefont {Kadau}, \citenamefont {Germann}, \citenamefont {Lomdahl},\ and\ \citenamefont {Holian}}]{10.1063/1.1483551}%
  \BibitemOpen
  \bibfield  {author} {\bibinfo {author} {\bibfnamefont {K.}~\bibnamefont {Kadau}}, \bibinfo {author} {\bibfnamefont {T.~C.}\ \bibnamefont {Germann}}, \bibinfo {author} {\bibfnamefont {P.~S.}\ \bibnamefont {Lomdahl}},\ and\ \bibinfo {author} {\bibfnamefont {B.~L.}\ \bibnamefont {Holian}},\ }\bibfield  {title} {\bibinfo {title} {Shock‐induced structural phase transformations studied by large‐scale molecular‐dynamics simulations},\ }\href {https://doi.org/10.1063/1.1483551} {\bibfield  {journal} {\bibinfo  {journal} {AIP Conference Proceedings}\ }\textbf {\bibinfo {volume} {620}},\ \bibinfo {pages} {351} (\bibinfo {year} {2002}{\natexlab{b}})}\BibitemShut {NoStop}%
\bibitem [{\citenamefont {Li}\ \emph {et~al.}(2019)\citenamefont {Li}, \citenamefont {Hahn}, \citenamefont {Yao}, \citenamefont {Germann},\ and\ \citenamefont {Zhang}}]{LI201951}%
  \BibitemOpen
  \bibfield  {author} {\bibinfo {author} {\bibfnamefont {W.}~\bibnamefont {Li}}, \bibinfo {author} {\bibfnamefont {E.~N.}\ \bibnamefont {Hahn}}, \bibinfo {author} {\bibfnamefont {X.}~\bibnamefont {Yao}}, \bibinfo {author} {\bibfnamefont {T.~C.}\ \bibnamefont {Germann}},\ and\ \bibinfo {author} {\bibfnamefont {X.}~\bibnamefont {Zhang}},\ }\bibfield  {title} {\bibinfo {title} {Shock induced damage and fracture in sic at elevated temperature and high strain rate},\ }\href {https://doi.org/https://doi.org/10.1016/j.actamat.2018.12.035} {\bibfield  {journal} {\bibinfo  {journal} {Acta Materialia}\ }\textbf {\bibinfo {volume} {167}},\ \bibinfo {pages} {51} (\bibinfo {year} {2019})}\BibitemShut {NoStop}%
\bibitem [{\citenamefont {Dremov}\ \emph {et~al.}(2006)\citenamefont {Dremov}, \citenamefont {Petrovtsev}, \citenamefont {Sapozhnikov}, \citenamefont {Smirnova}, \citenamefont {Preston},\ and\ \citenamefont {Zocher}}]{PhysRevB.74.144110}%
  \BibitemOpen
  \bibfield  {author} {\bibinfo {author} {\bibfnamefont {V.}~\bibnamefont {Dremov}}, \bibinfo {author} {\bibfnamefont {A.}~\bibnamefont {Petrovtsev}}, \bibinfo {author} {\bibfnamefont {P.}~\bibnamefont {Sapozhnikov}}, \bibinfo {author} {\bibfnamefont {M.}~\bibnamefont {Smirnova}}, \bibinfo {author} {\bibfnamefont {D.~L.}\ \bibnamefont {Preston}},\ and\ \bibinfo {author} {\bibfnamefont {M.~A.}\ \bibnamefont {Zocher}},\ }\bibfield  {title} {\bibinfo {title} {Molecular dynamics simulations of the initial stages of spall in nanocrystalline copper},\ }\href {https://doi.org/10.1103/PhysRevB.74.144110} {\bibfield  {journal} {\bibinfo  {journal} {Phys. Rev. B}\ }\textbf {\bibinfo {volume} {74}},\ \bibinfo {pages} {144110} (\bibinfo {year} {2006})}\BibitemShut {NoStop}%
\bibitem [{\citenamefont {Durand}\ \emph {et~al.}(2020)\citenamefont {Durand}, \citenamefont {Soulard}, \citenamefont {Colombet},\ and\ \citenamefont {Prat}}]{ref6DandS-microjetting}%
  \BibitemOpen
  \bibfield  {author} {\bibinfo {author} {\bibfnamefont {O.}~\bibnamefont {Durand}}, \bibinfo {author} {\bibfnamefont {L.}~\bibnamefont {Soulard}}, \bibinfo {author} {\bibfnamefont {L.}~\bibnamefont {Colombet}},\ and\ \bibinfo {author} {\bibfnamefont {R.}~\bibnamefont {Prat}},\ }\bibfield  {title} {\bibinfo {title} {Influence of the phase transitions of shock-loaded tin on microjetting and ejecta production using molecular dynamics simulations},\ }\href@noop {} {\bibfield  {journal} {\bibinfo  {journal} {Journal of Applied Physics}\ }\textbf {\bibinfo {volume} {127}} (\bibinfo {year} {2020})}\BibitemShut {NoStop}%
\bibitem [{\citenamefont {Liao}\ \emph {et~al.}(2014)\citenamefont {Liao}, \citenamefont {Xiang}, \citenamefont {Zeng},\ and\ \citenamefont {Chen}}]{ref4YiLiao}%
  \BibitemOpen
  \bibfield  {author} {\bibinfo {author} {\bibfnamefont {Y.}~\bibnamefont {Liao}}, \bibinfo {author} {\bibfnamefont {M.}~\bibnamefont {Xiang}}, \bibinfo {author} {\bibfnamefont {X.}~\bibnamefont {Zeng}},\ and\ \bibinfo {author} {\bibfnamefont {J.}~\bibnamefont {Chen}},\ }\bibfield  {title} {\bibinfo {title} {Molecular dynamics study of the micro-spallation of single crystal tin},\ }\href {https://doi.org/https://doi.org/10.1016/j.commatsci.2014.07.014} {\bibfield  {journal} {\bibinfo  {journal} {Computational Materials Science}\ }\textbf {\bibinfo {volume} {95}},\ \bibinfo {pages} {89} (\bibinfo {year} {2014})}\BibitemShut {NoStop}%
\bibitem [{\citenamefont {Wang}\ \emph {et~al.}(2021)\citenamefont {Wang}, \citenamefont {He}, \citenamefont {Zhou},\ and\ \citenamefont {Wang}}]{ref5Xin-XinWang}%
  \BibitemOpen
  \bibfield  {author} {\bibinfo {author} {\bibfnamefont {X.-X.}\ \bibnamefont {Wang}}, \bibinfo {author} {\bibfnamefont {A.-M.}\ \bibnamefont {He}}, \bibinfo {author} {\bibfnamefont {T.-T.}\ \bibnamefont {Zhou}},\ and\ \bibinfo {author} {\bibfnamefont {P.}~\bibnamefont {Wang}},\ }\bibfield  {title} {\bibinfo {title} {Spall damage in single crystal tin under shock wave loading: A molecular dynamics simulation},\ }\href {https://doi.org/https://doi.org/10.1016/j.mechmat.2021.103991} {\bibfield  {journal} {\bibinfo  {journal} {Mechanics of Materials}\ }\textbf {\bibinfo {volume} {160}},\ \bibinfo {pages} {103991} (\bibinfo {year} {2021})}\BibitemShut {NoStop}%
\bibitem [{\citenamefont {Wu}\ \emph {et~al.}(2018)\citenamefont {Wu}, \citenamefont {Wu}, \citenamefont {Zhu}, \citenamefont {Wang}, \citenamefont {He},\ and\ \citenamefont {Wu}}]{wu2018molecular}%
  \BibitemOpen
  \bibfield  {author} {\bibinfo {author} {\bibfnamefont {B.}~\bibnamefont {Wu}}, \bibinfo {author} {\bibfnamefont {F.}~\bibnamefont {Wu}}, \bibinfo {author} {\bibfnamefont {Y.}~\bibnamefont {Zhu}}, \bibinfo {author} {\bibfnamefont {P.}~\bibnamefont {Wang}}, \bibinfo {author} {\bibfnamefont {A.}~\bibnamefont {He}},\ and\ \bibinfo {author} {\bibfnamefont {H.}~\bibnamefont {Wu}},\ }\bibfield  {title} {\bibinfo {title} {Molecular dynamics simulations of ejecta production from sinusoidal tin surfaces under supported and unsupported shocks},\ }\href@noop {} {\bibfield  {journal} {\bibinfo  {journal} {AIP Advances}\ }\textbf {\bibinfo {volume} {8}} (\bibinfo {year} {2018})}\BibitemShut {NoStop}%
\bibitem [{\citenamefont {Stukowski}(2012)}]{stukowski2012structure}%
  \BibitemOpen
  \bibfield  {author} {\bibinfo {author} {\bibfnamefont {A.}~\bibnamefont {Stukowski}},\ }\bibfield  {title} {\bibinfo {title} {Structure identification methods for atomistic simulations of crystalline materials},\ }\href@noop {} {\bibfield  {journal} {\bibinfo  {journal} {Modelling and Simulation in Materials Science and Engineering}\ }\textbf {\bibinfo {volume} {20}},\ \bibinfo {pages} {045021} (\bibinfo {year} {2012})}\BibitemShut {NoStop}%
\bibitem [{\citenamefont {Baskes}\ \emph {et~al.}(1989)\citenamefont {Baskes}, \citenamefont {Nelson},\ and\ \citenamefont {Wright}}]{ref9meam}%
  \BibitemOpen
  \bibfield  {author} {\bibinfo {author} {\bibfnamefont {M.~I.}\ \bibnamefont {Baskes}}, \bibinfo {author} {\bibfnamefont {J.~S.}\ \bibnamefont {Nelson}},\ and\ \bibinfo {author} {\bibfnamefont {A.~F.}\ \bibnamefont {Wright}},\ }\bibfield  {title} {\bibinfo {title} {Semiempirical modified embedded-atom potentials for silicon and germanium},\ }\href {https://doi.org/10.1103/PhysRevB.40.6085} {\bibfield  {journal} {\bibinfo  {journal} {Phys. Rev. B}\ }\textbf {\bibinfo {volume} {40}},\ \bibinfo {pages} {6085} (\bibinfo {year} {1989})}\BibitemShut {NoStop}%
\bibitem [{\citenamefont {Ravelo}\ and\ \citenamefont {Baskes}(1997)}]{ref10meam1997}%
  \BibitemOpen
  \bibfield  {author} {\bibinfo {author} {\bibfnamefont {R.}~\bibnamefont {Ravelo}}\ and\ \bibinfo {author} {\bibfnamefont {M.}~\bibnamefont {Baskes}},\ }\bibfield  {title} {\bibinfo {title} {Equilibrium and thermodynamic properties of grey, white, and liquid tin},\ }\href {https://doi.org/10.1103/PhysRevLett.79.2482} {\bibfield  {journal} {\bibinfo  {journal} {Phys. Rev. Lett.}\ }\textbf {\bibinfo {volume} {79}},\ \bibinfo {pages} {2482} (\bibinfo {year} {1997})}\BibitemShut {NoStop}%
\bibitem [{\citenamefont {Daw}\ and\ \citenamefont {Baskes}(1984)}]{ref7eam}%
  \BibitemOpen
  \bibfield  {author} {\bibinfo {author} {\bibfnamefont {M.~S.}\ \bibnamefont {Daw}}\ and\ \bibinfo {author} {\bibfnamefont {M.~I.}\ \bibnamefont {Baskes}},\ }\bibfield  {title} {\bibinfo {title} {Embedded-atom method: Derivation and application to impurities, surfaces, and other defects in metals},\ }\href {https://doi.org/10.1103/PhysRevB.29.6443} {\bibfield  {journal} {\bibinfo  {journal} {Phys. Rev. B}\ }\textbf {\bibinfo {volume} {29}},\ \bibinfo {pages} {6443} (\bibinfo {year} {1984})}\BibitemShut {NoStop}%
\bibitem [{\citenamefont {Sapozhnikov}\ \emph {et~al.}(2014)\citenamefont {Sapozhnikov}, \citenamefont {Ionov}, \citenamefont {Dremov}, \citenamefont {Soulard},\ and\ \citenamefont {Durand}}]{ref8eam2014}%
  \BibitemOpen
  \bibfield  {author} {\bibinfo {author} {\bibfnamefont {F.}~\bibnamefont {Sapozhnikov}}, \bibinfo {author} {\bibfnamefont {G.}~\bibnamefont {Ionov}}, \bibinfo {author} {\bibfnamefont {V.}~\bibnamefont {Dremov}}, \bibinfo {author} {\bibfnamefont {L.}~\bibnamefont {Soulard}},\ and\ \bibinfo {author} {\bibfnamefont {O.}~\bibnamefont {Durand}},\ }\href {https://doi.org/10.1088/1742-6596/500/3/032017} {\bibfield  {journal} {\bibinfo  {journal} {J. Phys.: Conf. Ser.}\ }\textbf {\bibinfo {volume} {500}},\ \bibinfo {pages} {032017} (\bibinfo {year} {2014})}\BibitemShut {NoStop}%
\bibitem [{\citenamefont {Ko}\ \emph {et~al.}(2018)\citenamefont {Ko}, \citenamefont {Kim}, \citenamefont {Kwon},\ and\ \citenamefont {Lee}}]{ref11meam2018Ko}%
  \BibitemOpen
  \bibfield  {author} {\bibinfo {author} {\bibfnamefont {W.-S.}\ \bibnamefont {Ko}}, \bibinfo {author} {\bibfnamefont {D.-H.}\ \bibnamefont {Kim}}, \bibinfo {author} {\bibfnamefont {Y.-J.}\ \bibnamefont {Kwon}},\ and\ \bibinfo {author} {\bibfnamefont {M.~H.}\ \bibnamefont {Lee}},\ }\bibfield  {title} {\bibinfo {title} {Atomistic simulations of pure tin based on a new modified embedded-atom method interatomic potential},\ }\bibfield  {journal} {\bibinfo  {journal} {Metals}\ }\textbf {\bibinfo {volume} {8}},\ \href {https://doi.org/10.3390/met8110900} {10.3390/met8110900} (\bibinfo {year} {2018})\BibitemShut {NoStop}%
\bibitem [{\citenamefont {Behler}\ and\ \citenamefont {Parrinello}(2007)}]{PhysRevLett.98.146401}%
  \BibitemOpen
  \bibfield  {author} {\bibinfo {author} {\bibfnamefont {J.}~\bibnamefont {Behler}}\ and\ \bibinfo {author} {\bibfnamefont {M.}~\bibnamefont {Parrinello}},\ }\bibfield  {title} {\bibinfo {title} {Generalized neural-network representation of high-dimensional potential-energy surfaces},\ }\href {https://doi.org/10.1103/PhysRevLett.98.146401} {\bibfield  {journal} {\bibinfo  {journal} {Phys. Rev. Lett.}\ }\textbf {\bibinfo {volume} {98}},\ \bibinfo {pages} {146401} (\bibinfo {year} {2007})}\BibitemShut {NoStop}%
\bibitem [{\citenamefont {Bart\'ok}\ \emph {et~al.}(2010)\citenamefont {Bart\'ok}, \citenamefont {Payne}, \citenamefont {Kondor},\ and\ \citenamefont {Cs\'anyi}}]{PhysRevLett.104.136403}%
  \BibitemOpen
  \bibfield  {author} {\bibinfo {author} {\bibfnamefont {A.~P.}\ \bibnamefont {Bart\'ok}}, \bibinfo {author} {\bibfnamefont {M.~C.}\ \bibnamefont {Payne}}, \bibinfo {author} {\bibfnamefont {R.}~\bibnamefont {Kondor}},\ and\ \bibinfo {author} {\bibfnamefont {G.}~\bibnamefont {Cs\'anyi}},\ }\bibfield  {title} {\bibinfo {title} {Gaussian approximation potentials: The accuracy of quantum mechanics, without the electrons},\ }\href {https://doi.org/10.1103/PhysRevLett.104.136403} {\bibfield  {journal} {\bibinfo  {journal} {Phys. Rev. Lett.}\ }\textbf {\bibinfo {volume} {104}},\ \bibinfo {pages} {136403} (\bibinfo {year} {2010})}\BibitemShut {NoStop}%
\bibitem [{\citenamefont {Montavon}\ \emph {et~al.}(2013)\citenamefont {Montavon}, \citenamefont {Rupp}, \citenamefont {Gobre}, \citenamefont {Vazquez-Mayagoitia}, \citenamefont {Hansen}, \citenamefont {Tkatchenko}, \citenamefont {Müller},\ and\ \citenamefont {von Lilienfeld}}]{Montavon_2013}%
  \BibitemOpen
  \bibfield  {author} {\bibinfo {author} {\bibfnamefont {G.}~\bibnamefont {Montavon}}, \bibinfo {author} {\bibfnamefont {M.}~\bibnamefont {Rupp}}, \bibinfo {author} {\bibfnamefont {V.}~\bibnamefont {Gobre}}, \bibinfo {author} {\bibfnamefont {A.}~\bibnamefont {Vazquez-Mayagoitia}}, \bibinfo {author} {\bibfnamefont {K.}~\bibnamefont {Hansen}}, \bibinfo {author} {\bibfnamefont {A.}~\bibnamefont {Tkatchenko}}, \bibinfo {author} {\bibfnamefont {K.-R.}\ \bibnamefont {Müller}},\ and\ \bibinfo {author} {\bibfnamefont {O.~A.}\ \bibnamefont {von Lilienfeld}},\ }\bibfield  {title} {\bibinfo {title} {Machine learning of molecular electronic properties in chemical compound space},\ }\href {https://doi.org/10.1088/1367-2630/15/9/095003} {\bibfield  {journal} {\bibinfo  {journal} {New Journal of Physics}\ }\textbf {\bibinfo {volume} {15}},\ \bibinfo {pages} {095003} (\bibinfo {year} {2013})}\BibitemShut {NoStop}%
\bibitem [{\citenamefont {Schütt}\ \emph {et~al.}(2017)\citenamefont {Schütt}, \citenamefont {Arbabzadah}, \citenamefont {Chmiela}, \citenamefont {Müller},\ and\ \citenamefont {Tkatchenko}}]{Schütt2017}%
  \BibitemOpen
  \bibfield  {author} {\bibinfo {author} {\bibfnamefont {K.~T.}\ \bibnamefont {Schütt}}, \bibinfo {author} {\bibfnamefont {F.}~\bibnamefont {Arbabzadah}}, \bibinfo {author} {\bibfnamefont {S.}~\bibnamefont {Chmiela}}, \bibinfo {author} {\bibfnamefont {K.~R.}\ \bibnamefont {Müller}},\ and\ \bibinfo {author} {\bibfnamefont {A.}~\bibnamefont {Tkatchenko}},\ }\bibfield  {title} {\bibinfo {title} {Quantum-chemical insights from deep tensor neural networks},\ }\href {https://doi.org/10.1038/ncomms13890} {\bibfield  {journal} {\bibinfo  {journal} {Nature Communications}\ }\textbf {\bibinfo {volume} {8}},\ \bibinfo {pages} {13890} (\bibinfo {year} {2017})}\BibitemShut {NoStop}%
\bibitem [{\citenamefont {Zhang}\ \emph {et~al.}(2018{\natexlab{a}})\citenamefont {Zhang}, \citenamefont {Han}, \citenamefont {Wang}, \citenamefont {Car},\ and\ \citenamefont {E}}]{PhysRevLett.120.143001}%
  \BibitemOpen
  \bibfield  {author} {\bibinfo {author} {\bibfnamefont {L.}~\bibnamefont {Zhang}}, \bibinfo {author} {\bibfnamefont {J.}~\bibnamefont {Han}}, \bibinfo {author} {\bibfnamefont {H.}~\bibnamefont {Wang}}, \bibinfo {author} {\bibfnamefont {R.}~\bibnamefont {Car}},\ and\ \bibinfo {author} {\bibfnamefont {W.}~\bibnamefont {E}},\ }\bibfield  {title} {\bibinfo {title} {Deep potential molecular dynamics: A scalable model with the accuracy of quantum mechanics},\ }\href {https://doi.org/10.1103/PhysRevLett.120.143001} {\bibfield  {journal} {\bibinfo  {journal} {Phys. Rev. Lett.}\ }\textbf {\bibinfo {volume} {120}},\ \bibinfo {pages} {143001} (\bibinfo {year} {2018}{\natexlab{a}})}\BibitemShut {NoStop}%
\bibitem [{\citenamefont {Chen}\ \emph {et~al.}(2023)\citenamefont {Chen}, \citenamefont {Yuan}, \citenamefont {Liu}, \citenamefont {Geng}, \citenamefont {Zhang}, \citenamefont {Wang},\ and\ \citenamefont {Chen}}]{ref1DP-SCAN}%
  \BibitemOpen
  \bibfield  {author} {\bibinfo {author} {\bibfnamefont {T.}~\bibnamefont {Chen}}, \bibinfo {author} {\bibfnamefont {F.}~\bibnamefont {Yuan}}, \bibinfo {author} {\bibfnamefont {J.}~\bibnamefont {Liu}}, \bibinfo {author} {\bibfnamefont {H.}~\bibnamefont {Geng}}, \bibinfo {author} {\bibfnamefont {L.}~\bibnamefont {Zhang}}, \bibinfo {author} {\bibfnamefont {H.}~\bibnamefont {Wang}},\ and\ \bibinfo {author} {\bibfnamefont {M.}~\bibnamefont {Chen}},\ }\bibfield  {title} {\bibinfo {title} {Modeling the high-pressure solid and liquid phases of tin from deep potentials with ab initio accuracy},\ }\href {https://doi.org/10.1103/PhysRevMaterials.7.053603} {\bibfield  {journal} {\bibinfo  {journal} {Phys. Rev. Mater.}\ }\textbf {\bibinfo {volume} {7}},\ \bibinfo {pages} {053603} (\bibinfo {year} {2023})}\BibitemShut {NoStop}%
\bibitem [{\citenamefont {Nitol}\ \emph {et~al.}(2023)\citenamefont {Nitol}, \citenamefont {Dang}, \citenamefont {Fensin}, \citenamefont {Baskes}, \citenamefont {Dickel},\ and\ \citenamefont {Barrett}}]{ref12EAM-R}%
  \BibitemOpen
  \bibfield  {author} {\bibinfo {author} {\bibfnamefont {M.~S.}\ \bibnamefont {Nitol}}, \bibinfo {author} {\bibfnamefont {K.}~\bibnamefont {Dang}}, \bibinfo {author} {\bibfnamefont {S.~J.}\ \bibnamefont {Fensin}}, \bibinfo {author} {\bibfnamefont {M.~I.}\ \bibnamefont {Baskes}}, \bibinfo {author} {\bibfnamefont {D.~E.}\ \bibnamefont {Dickel}},\ and\ \bibinfo {author} {\bibfnamefont {C.~D.}\ \bibnamefont {Barrett}},\ }\bibfield  {title} {\bibinfo {title} {Hybrid interatomic potential for sn},\ }\href {https://doi.org/10.1103/PhysRevMaterials.7.043601} {\bibfield  {journal} {\bibinfo  {journal} {Phys. Rev. Mater.}\ }\textbf {\bibinfo {volume} {7}},\ \bibinfo {pages} {043601} (\bibinfo {year} {2023})}\BibitemShut {NoStop}%
\bibitem [{\citenamefont {Hohenberg}\ and\ \citenamefont {Kohn}(1964{\natexlab{a}})}]{PhysRev.136.B864}%
  \BibitemOpen
  \bibfield  {author} {\bibinfo {author} {\bibfnamefont {P.}~\bibnamefont {Hohenberg}}\ and\ \bibinfo {author} {\bibfnamefont {W.}~\bibnamefont {Kohn}},\ }\bibfield  {title} {\bibinfo {title} {Inhomogeneous electron gas},\ }\href {https://doi.org/10.1103/PhysRev.136.B864} {\bibfield  {journal} {\bibinfo  {journal} {Phys. Rev.}\ }\textbf {\bibinfo {volume} {136}},\ \bibinfo {pages} {B864} (\bibinfo {year} {1964}{\natexlab{a}})}\BibitemShut {NoStop}%
\bibitem [{\citenamefont {Kohn}\ and\ \citenamefont {Sham}(1965{\natexlab{a}})}]{PhysRev.140.A1133}%
  \BibitemOpen
  \bibfield  {author} {\bibinfo {author} {\bibfnamefont {W.}~\bibnamefont {Kohn}}\ and\ \bibinfo {author} {\bibfnamefont {L.~J.}\ \bibnamefont {Sham}},\ }\bibfield  {title} {\bibinfo {title} {Self-consistent equations including exchange and correlation effects},\ }\href {https://doi.org/10.1103/PhysRev.140.A1133} {\bibfield  {journal} {\bibinfo  {journal} {Phys. Rev.}\ }\textbf {\bibinfo {volume} {140}},\ \bibinfo {pages} {A1133} (\bibinfo {year} {1965}{\natexlab{a}})}\BibitemShut {NoStop}%
\bibitem [{\citenamefont {Sun}\ \emph {et~al.}(2015)\citenamefont {Sun}, \citenamefont {Ruzsinszky},\ and\ \citenamefont {Perdew}}]{ref24SCANfunc}%
  \BibitemOpen
  \bibfield  {author} {\bibinfo {author} {\bibfnamefont {J.}~\bibnamefont {Sun}}, \bibinfo {author} {\bibfnamefont {A.}~\bibnamefont {Ruzsinszky}},\ and\ \bibinfo {author} {\bibfnamefont {J.~P.}\ \bibnamefont {Perdew}},\ }\bibfield  {title} {\bibinfo {title} {Strongly constrained and appropriately normed semilocal density functional},\ }\href {https://doi.org/10.1103/PhysRevLett.115.036402} {\bibfield  {journal} {\bibinfo  {journal} {Phys. Rev. Lett.}\ }\textbf {\bibinfo {volume} {115}},\ \bibinfo {pages} {036402} (\bibinfo {year} {2015})}\BibitemShut {NoStop}%
\bibitem [{\citenamefont {Perdew}\ \emph {et~al.}(1996)\citenamefont {Perdew}, \citenamefont {Burke},\ and\ \citenamefont {Ernzerhof}}]{PBE}%
  \BibitemOpen
  \bibfield  {author} {\bibinfo {author} {\bibfnamefont {J.~P.}\ \bibnamefont {Perdew}}, \bibinfo {author} {\bibfnamefont {K.}~\bibnamefont {Burke}},\ and\ \bibinfo {author} {\bibfnamefont {M.}~\bibnamefont {Ernzerhof}},\ }\bibfield  {title} {\bibinfo {title} {Generalized gradient approximation made simple},\ }\href {https://doi.org/10.1103/PhysRevLett.77.3865} {\bibfield  {journal} {\bibinfo  {journal} {Phys. Rev. Lett.}\ }\textbf {\bibinfo {volume} {77}},\ \bibinfo {pages} {3865} (\bibinfo {year} {1996})}\BibitemShut {NoStop}%
\bibitem [{\citenamefont {Rehn}\ \emph {et~al.}(2021)\citenamefont {Rehn}, \citenamefont {Greeff}, \citenamefont {Burakovsky}, \citenamefont {Sheppard},\ and\ \citenamefont {Crockett}}]{PhysRevB.103.184102}%
  \BibitemOpen
  \bibfield  {author} {\bibinfo {author} {\bibfnamefont {D.~A.}\ \bibnamefont {Rehn}}, \bibinfo {author} {\bibfnamefont {C.~W.}\ \bibnamefont {Greeff}}, \bibinfo {author} {\bibfnamefont {L.}~\bibnamefont {Burakovsky}}, \bibinfo {author} {\bibfnamefont {D.~G.}\ \bibnamefont {Sheppard}},\ and\ \bibinfo {author} {\bibfnamefont {S.~D.}\ \bibnamefont {Crockett}},\ }\bibfield  {title} {\bibinfo {title} {Multiphase tin equation of state using density functional theory},\ }\href {https://doi.org/10.1103/PhysRevB.103.184102} {\bibfield  {journal} {\bibinfo  {journal} {Phys. Rev. B}\ }\textbf {\bibinfo {volume} {103}},\ \bibinfo {pages} {184102} (\bibinfo {year} {2021})}\BibitemShut {NoStop}%
\bibitem [{\citenamefont {Marsh}(1980)}]{ref13HuExp}%
  \BibitemOpen
  \bibfield  {author} {\bibinfo {author} {\bibfnamefont {S.~P.}\ \bibnamefont {Marsh}},\ }\href@noop {} {\emph {\bibinfo {title} {LASL shock Hugoniot data}}},\ Vol.~\bibinfo {volume} {5}\ (\bibinfo  {publisher} {Univ of California Press},\ \bibinfo {year} {1980})\BibitemShut {NoStop}%
\bibitem [{\citenamefont {De~Ress{\'e}guier}\ \emph {et~al.}(2014)\citenamefont {De~Ress{\'e}guier}, \citenamefont {Lescoute}, \citenamefont {Sollier}, \citenamefont {Prudhomme},\ and\ \citenamefont {Mercier}}]{de2014microjetting}%
  \BibitemOpen
  \bibfield  {author} {\bibinfo {author} {\bibfnamefont {T.}~\bibnamefont {De~Ress{\'e}guier}}, \bibinfo {author} {\bibfnamefont {E.}~\bibnamefont {Lescoute}}, \bibinfo {author} {\bibfnamefont {A.}~\bibnamefont {Sollier}}, \bibinfo {author} {\bibfnamefont {G.}~\bibnamefont {Prudhomme}},\ and\ \bibinfo {author} {\bibfnamefont {P.}~\bibnamefont {Mercier}},\ }\bibfield  {title} {\bibinfo {title} {Microjetting from grooved surfaces in metallic samples subjected to laser driven shocks},\ }\href@noop {} {\bibfield  {journal} {\bibinfo  {journal} {Journal of Applied Physics}\ }\textbf {\bibinfo {volume} {115}} (\bibinfo {year} {2014})}\BibitemShut {NoStop}%
\bibitem [{\citenamefont {Saunders}\ \emph {et~al.}(2024)\citenamefont {Saunders}, \citenamefont {Sun}, \citenamefont {Horwitz}, \citenamefont {Ali}, \citenamefont {Eggert}, \citenamefont {Mackay}, \citenamefont {Morgan}, \citenamefont {Najjar}, \citenamefont {Park}, \citenamefont {Ping},\ and\ \citenamefont {Pino}}]{Saunders2024}%
  \BibitemOpen
  \bibfield  {author} {\bibinfo {author} {\bibfnamefont {A.~M.}\ \bibnamefont {Saunders}}, \bibinfo {author} {\bibfnamefont {Y.}~\bibnamefont {Sun}}, \bibinfo {author} {\bibfnamefont {J.~A.~K.}\ \bibnamefont {Horwitz}}, \bibinfo {author} {\bibfnamefont {S.~J.}\ \bibnamefont {Ali}}, \bibinfo {author} {\bibfnamefont {J.~H.}\ \bibnamefont {Eggert}}, \bibinfo {author} {\bibfnamefont {K.~K.}\ \bibnamefont {Mackay}}, \bibinfo {author} {\bibfnamefont {B.~E.}\ \bibnamefont {Morgan}}, \bibinfo {author} {\bibfnamefont {F.~M.}\ \bibnamefont {Najjar}}, \bibinfo {author} {\bibfnamefont {H.-S.}\ \bibnamefont {Park}}, \bibinfo {author} {\bibfnamefont {Y.}~\bibnamefont {Ping}},\ and\ \bibinfo {author} {\bibfnamefont {J.}~\bibnamefont {Pino}},\ }\bibfield  {title} {\bibinfo {title} {Interactions of laser-driven tin ejecta microjets over phase transition boundaries},\ }\href {https://doi.org/10.1063/5.0211440} {\bibfield  {journal} {\bibinfo  {journal} {Journal of Applied Physics}\ }\textbf {\bibinfo {volume} {136}},\ \bibinfo
  {pages} {025901} (\bibinfo {year} {2024})}\BibitemShut {NoStop}%
\bibitem [{\citenamefont {Yang}\ \emph {et~al.}(2022)\citenamefont {Yang}, \citenamefont {Zhao}, \citenamefont {Gao}, \citenamefont {Chen}, \citenamefont {Zeng},\ and\ \citenamefont {Wang}}]{10.1063/5.0099331}%
  \BibitemOpen
  \bibfield  {author} {\bibinfo {author} {\bibfnamefont {X.}~\bibnamefont {Yang}}, \bibinfo {author} {\bibfnamefont {H.}~\bibnamefont {Zhao}}, \bibinfo {author} {\bibfnamefont {X.}~\bibnamefont {Gao}}, \bibinfo {author} {\bibfnamefont {Z.}~\bibnamefont {Chen}}, \bibinfo {author} {\bibfnamefont {X.}~\bibnamefont {Zeng}},\ and\ \bibinfo {author} {\bibfnamefont {F.}~\bibnamefont {Wang}},\ }\bibfield  {title} {\bibinfo {title} {Molecular dynamics study on spallation fracture in single crystal and nanocrystalline tin},\ }\href {https://doi.org/10.1063/5.0099331} {\bibfield  {journal} {\bibinfo  {journal} {Journal of Applied Physics}\ }\textbf {\bibinfo {volume} {132}},\ \bibinfo {pages} {075903} (\bibinfo {year} {2022})}\BibitemShut {NoStop}%
\bibitem [{\citenamefont {Soulard}\ and\ \citenamefont {Durand}(2020)}]{10.1063/5.0003089}%
  \BibitemOpen
  \bibfield  {author} {\bibinfo {author} {\bibfnamefont {L.}~\bibnamefont {Soulard}}\ and\ \bibinfo {author} {\bibfnamefont {O.}~\bibnamefont {Durand}},\ }\bibfield  {title} {\bibinfo {title} {Observation of phase transitions in shocked tin by molecular dynamics},\ }\href {https://doi.org/10.1063/5.0003089} {\bibfield  {journal} {\bibinfo  {journal} {Journal of Applied Physics}\ }\textbf {\bibinfo {volume} {127}},\ \bibinfo {pages} {165901} (\bibinfo {year} {2020})}\BibitemShut {NoStop}%
\bibitem [{\citenamefont {Needham}(2018)}]{Needham2018}%
  \BibitemOpen
  \bibfield  {author} {\bibinfo {author} {\bibfnamefont {C.~E.}\ \bibnamefont {Needham}},\ }\bibinfo {title} {The rankine-hugoniot relations},\ in\ \href {https://doi.org/10.1007/978-3-319-65382-2_3} {\emph {\bibinfo {booktitle} {Blast Waves}}}\ (\bibinfo  {publisher} {Springer International Publishing},\ \bibinfo {address} {Cham},\ \bibinfo {year} {2018})\ pp.\ \bibinfo {pages} {9--17}\BibitemShut {NoStop}%
\bibitem [{\citenamefont {Jiang}\ \emph {et~al.}(2021)\citenamefont {Jiang}, \citenamefont {Zhang}, \citenamefont {Zhang},\ and\ \citenamefont {Wang}}]{Jiang_2021}%
  \BibitemOpen
  \bibfield  {author} {\bibinfo {author} {\bibfnamefont {W.}~\bibnamefont {Jiang}}, \bibinfo {author} {\bibfnamefont {Y.}~\bibnamefont {Zhang}}, \bibinfo {author} {\bibfnamefont {L.}~\bibnamefont {Zhang}},\ and\ \bibinfo {author} {\bibfnamefont {H.}~\bibnamefont {Wang}},\ }\bibfield  {title} {\bibinfo {title} {Accurate deep potential model for the al–cu–mg alloy in the full concentration space*},\ }\href {https://doi.org/10.1088/1674-1056/abf134} {\bibfield  {journal} {\bibinfo  {journal} {Chinese Physics B}\ }\textbf {\bibinfo {volume} {30}},\ \bibinfo {pages} {050706} (\bibinfo {year} {2021})}\BibitemShut {NoStop}%
\bibitem [{\citenamefont {Niu}\ \emph {et~al.}(2020)\citenamefont {Niu}, \citenamefont {Piaggi},\ and\ \citenamefont {Parrinello}}]{NC2020}%
  \BibitemOpen
  \bibfield  {author} {\bibinfo {author} {\bibfnamefont {H.}~\bibnamefont {Niu}}, \bibinfo {author} {\bibfnamefont {L.~B. P.~M.}\ \bibnamefont {Piaggi}},\ and\ \bibinfo {author} {\bibfnamefont {M.}~\bibnamefont {Parrinello}},\ }\bibfield  {title} {\bibinfo {title} {Ab initio phase diagram and nucleation of gallium},\ }\href {https://doi.org/10.1038/s41467-020-163729} {\bibfield  {journal} {\bibinfo  {journal} {Nature Communications}\ }\textbf {\bibinfo {volume} {11}},\ \bibinfo {pages} {2654} (\bibinfo {year} {2020})}\BibitemShut {NoStop}%
\bibitem [{\citenamefont {Zhang}\ \emph {et~al.}(2021)\citenamefont {Zhang}, \citenamefont {Wang}, \citenamefont {Car},\ and\ \citenamefont {E}}]{PhysRevLett.126.236001}%
  \BibitemOpen
  \bibfield  {author} {\bibinfo {author} {\bibfnamefont {L.}~\bibnamefont {Zhang}}, \bibinfo {author} {\bibfnamefont {H.}~\bibnamefont {Wang}}, \bibinfo {author} {\bibfnamefont {R.}~\bibnamefont {Car}},\ and\ \bibinfo {author} {\bibfnamefont {W.}~\bibnamefont {E}},\ }\bibfield  {title} {\bibinfo {title} {Phase diagram of a deep potential water model},\ }\href {https://doi.org/10.1103/PhysRevLett.126.236001} {\bibfield  {journal} {\bibinfo  {journal} {Phys. Rev. Lett.}\ }\textbf {\bibinfo {volume} {126}},\ \bibinfo {pages} {236001} (\bibinfo {year} {2021})}\BibitemShut {NoStop}%
\bibitem [{\citenamefont {Xu}\ \emph {et~al.}(2020)\citenamefont {Xu}, \citenamefont {Zhang}, \citenamefont {Zhang}, \citenamefont {Chen}, \citenamefont {Santra},\ and\ \citenamefont {Wu}}]{PhysRevB.102.214113}%
  \BibitemOpen
  \bibfield  {author} {\bibinfo {author} {\bibfnamefont {J.}~\bibnamefont {Xu}}, \bibinfo {author} {\bibfnamefont {C.}~\bibnamefont {Zhang}}, \bibinfo {author} {\bibfnamefont {L.}~\bibnamefont {Zhang}}, \bibinfo {author} {\bibfnamefont {M.}~\bibnamefont {Chen}}, \bibinfo {author} {\bibfnamefont {B.}~\bibnamefont {Santra}},\ and\ \bibinfo {author} {\bibfnamefont {X.}~\bibnamefont {Wu}},\ }\bibfield  {title} {\bibinfo {title} {Isotope effects in molecular structures and electronic properties of liquid water via deep potential molecular dynamics based on the scan functional},\ }\href {https://doi.org/10.1103/PhysRevB.102.214113} {\bibfield  {journal} {\bibinfo  {journal} {Phys. Rev. B}\ }\textbf {\bibinfo {volume} {102}},\ \bibinfo {pages} {214113} (\bibinfo {year} {2020})}\BibitemShut {NoStop}%
\bibitem [{\citenamefont {Liu}\ \emph {et~al.}(2020)\citenamefont {Liu}, \citenamefont {Lu},\ and\ \citenamefont {Chen}}]{Liu_2020}%
  \BibitemOpen
  \bibfield  {author} {\bibinfo {author} {\bibfnamefont {Q.}~\bibnamefont {Liu}}, \bibinfo {author} {\bibfnamefont {D.}~\bibnamefont {Lu}},\ and\ \bibinfo {author} {\bibfnamefont {M.}~\bibnamefont {Chen}},\ }\bibfield  {title} {\bibinfo {title} {Structure and dynamics of warm dense aluminum: a molecular dynamics study with density functional theory and deep potential},\ }\href {https://doi.org/10.1088/1361-648X/ab5890} {\bibfield  {journal} {\bibinfo  {journal} {Journal of Physics: Condensed Matter}\ }\textbf {\bibinfo {volume} {32}},\ \bibinfo {pages} {144002} (\bibinfo {year} {2020})}\BibitemShut {NoStop}%
\bibitem [{\citenamefont {Liu}\ \emph {et~al.}(2021)\citenamefont {Liu}, \citenamefont {Li},\ and\ \citenamefont {Chen}}]{10.1063/5.0030123}%
  \BibitemOpen
  \bibfield  {author} {\bibinfo {author} {\bibfnamefont {Q.}~\bibnamefont {Liu}}, \bibinfo {author} {\bibfnamefont {J.}~\bibnamefont {Li}},\ and\ \bibinfo {author} {\bibfnamefont {M.}~\bibnamefont {Chen}},\ }\bibfield  {title} {\bibinfo {title} {Thermal transport by electrons and ions in warm dense aluminum: A combined density functional theory and deep potential study},\ }\href {https://doi.org/10.1063/5.0030123} {\bibfield  {journal} {\bibinfo  {journal} {Matter and Radiation at Extremes}\ }\textbf {\bibinfo {volume} {6}},\ \bibinfo {pages} {026902} (\bibinfo {year} {2021})}\BibitemShut {NoStop}%
\bibitem [{\citenamefont {Wang}\ \emph {et~al.}(2022)\citenamefont {Wang}, \citenamefont {Wang}, \citenamefont {Zhang}, \citenamefont {Dai},\ and\ \citenamefont {Wang}}]{Wang_2022}%
  \BibitemOpen
  \bibfield  {author} {\bibinfo {author} {\bibfnamefont {X.}~\bibnamefont {Wang}}, \bibinfo {author} {\bibfnamefont {Y.}~\bibnamefont {Wang}}, \bibinfo {author} {\bibfnamefont {L.}~\bibnamefont {Zhang}}, \bibinfo {author} {\bibfnamefont {F.}~\bibnamefont {Dai}},\ and\ \bibinfo {author} {\bibfnamefont {H.}~\bibnamefont {Wang}},\ }\bibfield  {title} {\bibinfo {title} {A tungsten deep neural-network potential for simulating mechanical property degradation under fusion service environment},\ }\href {https://doi.org/10.1088/1741-4326/ac888b} {\bibfield  {journal} {\bibinfo  {journal} {Nuclear Fusion}\ }\textbf {\bibinfo {volume} {62}},\ \bibinfo {pages} {126013} (\bibinfo {year} {2022})}\BibitemShut {NoStop}%
\bibitem [{\citenamefont {Wang}\ \emph {et~al.}(2023)\citenamefont {Wang}, \citenamefont {Wang}, \citenamefont {Xu}, \citenamefont {Dai}, \citenamefont {Liu}, \citenamefont {Lu},\ and\ \citenamefont {Wang}}]{PhysRevMaterials.7.093601}%
  \BibitemOpen
  \bibfield  {author} {\bibinfo {author} {\bibfnamefont {X.-Y.}\ \bibnamefont {Wang}}, \bibinfo {author} {\bibfnamefont {Y.-N.}\ \bibnamefont {Wang}}, \bibinfo {author} {\bibfnamefont {K.}~\bibnamefont {Xu}}, \bibinfo {author} {\bibfnamefont {F.-Z.}\ \bibnamefont {Dai}}, \bibinfo {author} {\bibfnamefont {H.-F.}\ \bibnamefont {Liu}}, \bibinfo {author} {\bibfnamefont {G.-H.}\ \bibnamefont {Lu}},\ and\ \bibinfo {author} {\bibfnamefont {H.}~\bibnamefont {Wang}},\ }\bibfield  {title} {\bibinfo {title} {Deep neural network potential for simulating hydrogen blistering in tungsten},\ }\href {https://doi.org/10.1103/PhysRevMaterials.7.093601} {\bibfield  {journal} {\bibinfo  {journal} {Phys. Rev. Mater.}\ }\textbf {\bibinfo {volume} {7}},\ \bibinfo {pages} {093601} (\bibinfo {year} {2023})}\BibitemShut {NoStop}%
\bibitem [{\citenamefont {Liu}\ \emph {et~al.}(2023)\citenamefont {Liu}, \citenamefont {Liu}, \citenamefont {Cao},\ and\ \citenamefont {Chen}}]{D2CP04105G}%
  \BibitemOpen
  \bibfield  {author} {\bibinfo {author} {\bibfnamefont {J.}~\bibnamefont {Liu}}, \bibinfo {author} {\bibfnamefont {R.}~\bibnamefont {Liu}}, \bibinfo {author} {\bibfnamefont {Y.}~\bibnamefont {Cao}},\ and\ \bibinfo {author} {\bibfnamefont {M.}~\bibnamefont {Chen}},\ }\bibfield  {title} {\bibinfo {title} {Solvation structures of calcium and magnesium ions in water with the presence of hydroxide: a study by deep potential molecular dynamics},\ }\href {https://doi.org/10.1039/D2CP04105G} {\bibfield  {journal} {\bibinfo  {journal} {Phys. Chem. Chem. Phys.}\ }\textbf {\bibinfo {volume} {25}},\ \bibinfo {pages} {983} (\bibinfo {year} {2023})}\BibitemShut {NoStop}%
\bibitem [{\citenamefont {Liu}\ \emph {et~al.}(2024)\citenamefont {Liu}, \citenamefont {Lu},\ and\ \citenamefont {Wang}}]{POD2024}%
  \BibitemOpen
  \bibfield  {author} {\bibinfo {author} {\bibfnamefont {Z.}~\bibnamefont {Liu}}, \bibinfo {author} {\bibfnamefont {A.-H.}\ \bibnamefont {Lu}},\ and\ \bibinfo {author} {\bibfnamefont {D.}~\bibnamefont {Wang}},\ }\bibfield  {title} {\bibinfo {title} {Deep potential molecular dynamics study of propane oxidative dehydrogenation},\ }\href {https://doi.org/10.1021/acs.jpca.3c07859} {\bibfield  {journal} {\bibinfo  {journal} {The Journal of Physical Chemistry A}\ }\textbf {\bibinfo {volume} {128}},\ \bibinfo {pages} {1656} (\bibinfo {year} {2024})}\BibitemShut {NoStop}%
\bibitem [{\citenamefont {Zhang}\ \emph {et~al.}(2020)\citenamefont {Zhang}, \citenamefont {Wang}, \citenamefont {Chen}, \citenamefont {Zeng}, \citenamefont {Zhang}, \citenamefont {Wang},\ and\ \citenamefont {E}}]{ref3dpgen}%
  \BibitemOpen
  \bibfield  {author} {\bibinfo {author} {\bibfnamefont {Y.}~\bibnamefont {Zhang}}, \bibinfo {author} {\bibfnamefont {H.}~\bibnamefont {Wang}}, \bibinfo {author} {\bibfnamefont {W.}~\bibnamefont {Chen}}, \bibinfo {author} {\bibfnamefont {J.}~\bibnamefont {Zeng}}, \bibinfo {author} {\bibfnamefont {L.}~\bibnamefont {Zhang}}, \bibinfo {author} {\bibfnamefont {H.}~\bibnamefont {Wang}},\ and\ \bibinfo {author} {\bibfnamefont {W.}~\bibnamefont {E}},\ }\bibfield  {title} {\bibinfo {title} {Dp-gen: A concurrent learning platform for the generation of reliable deep learning based potential energy models},\ }\bibfield  {journal} {\bibinfo  {journal} {Mendeley Data}\ }\textbf {\bibinfo {volume} {1}},\ \href {https://doi.org/10.17632/sxybkgc5xc.1} {10.17632/sxybkgc5xc.1} (\bibinfo {year} {2020})\BibitemShut {NoStop}%
\bibitem [{\citenamefont {Wang}\ \emph {et~al.}(2018)\citenamefont {Wang}, \citenamefont {Zhang}, \citenamefont {Han} \emph {et~al.}}]{wang2018deepmd}%
  \BibitemOpen
  \bibfield  {author} {\bibinfo {author} {\bibfnamefont {H.}~\bibnamefont {Wang}}, \bibinfo {author} {\bibfnamefont {L.}~\bibnamefont {Zhang}}, \bibinfo {author} {\bibfnamefont {J.}~\bibnamefont {Han}}, \emph {et~al.},\ }\bibfield  {title} {\bibinfo {title} {Deepmd-kit: A deep learning package for many-body potential energy representation and molecular dynamics},\ }\href@noop {} {\bibfield  {journal} {\bibinfo  {journal} {Computer Physics Communications}\ }\textbf {\bibinfo {volume} {228}},\ \bibinfo {pages} {178} (\bibinfo {year} {2018})}\BibitemShut {NoStop}%
\bibitem [{\citenamefont {Zhang}\ \emph {et~al.}(2018{\natexlab{b}})\citenamefont {Zhang}, \citenamefont {Han}, \citenamefont {Wang}, \citenamefont {Saidi}, \citenamefont {Car} \emph {et~al.}}]{ref2dp}%
  \BibitemOpen
  \bibfield  {author} {\bibinfo {author} {\bibfnamefont {L.}~\bibnamefont {Zhang}}, \bibinfo {author} {\bibfnamefont {J.}~\bibnamefont {Han}}, \bibinfo {author} {\bibfnamefont {H.}~\bibnamefont {Wang}}, \bibinfo {author} {\bibfnamefont {W.}~\bibnamefont {Saidi}}, \bibinfo {author} {\bibfnamefont {R.}~\bibnamefont {Car}}, \emph {et~al.},\ }\bibfield  {title} {\bibinfo {title} {End-to-end symmetry preserving inter-atomic potential energy model for finite and extended systems},\ }\href@noop {} {\bibfield  {journal} {\bibinfo  {journal} {Advances in neural information processing systems}\ }\textbf {\bibinfo {volume} {31}} (\bibinfo {year} {2018}{\natexlab{b}})}\BibitemShut {NoStop}%
\bibitem [{\citenamefont {Thompson}\ \emph {et~al.}(2022)\citenamefont {Thompson}, \citenamefont {Aktulga}, \citenamefont {Berger}, \citenamefont {Bolintineanu}, \citenamefont {Brown}, \citenamefont {Crozier}, \citenamefont {in~'t Veld}, \citenamefont {Kohlmeyer}, \citenamefont {Moore}, \citenamefont {Nguyen}, \citenamefont {Shan}, \citenamefont {Stevens}, \citenamefont {Tranchida}, \citenamefont {Trott},\ and\ \citenamefont {Plimpton}}]{ref18lammps}%
  \BibitemOpen
  \bibfield  {author} {\bibinfo {author} {\bibfnamefont {A.~P.}\ \bibnamefont {Thompson}}, \bibinfo {author} {\bibfnamefont {H.~M.}\ \bibnamefont {Aktulga}}, \bibinfo {author} {\bibfnamefont {R.}~\bibnamefont {Berger}}, \bibinfo {author} {\bibfnamefont {D.~S.}\ \bibnamefont {Bolintineanu}}, \bibinfo {author} {\bibfnamefont {W.~M.}\ \bibnamefont {Brown}}, \bibinfo {author} {\bibfnamefont {P.~S.}\ \bibnamefont {Crozier}}, \bibinfo {author} {\bibfnamefont {P.~J.}\ \bibnamefont {in~'t Veld}}, \bibinfo {author} {\bibfnamefont {A.}~\bibnamefont {Kohlmeyer}}, \bibinfo {author} {\bibfnamefont {S.~G.}\ \bibnamefont {Moore}}, \bibinfo {author} {\bibfnamefont {T.~D.}\ \bibnamefont {Nguyen}}, \bibinfo {author} {\bibfnamefont {R.}~\bibnamefont {Shan}}, \bibinfo {author} {\bibfnamefont {M.~J.}\ \bibnamefont {Stevens}}, \bibinfo {author} {\bibfnamefont {J.}~\bibnamefont {Tranchida}}, \bibinfo {author} {\bibfnamefont {C.}~\bibnamefont {Trott}},\ and\ \bibinfo {author} {\bibfnamefont {S.~J.}\ \bibnamefont {Plimpton}},\
  }\bibfield  {title} {\bibinfo {title} {{LAMMPS} - a flexible simulation tool for particle-based materials modeling at the atomic, meso, and continuum scales},\ }\href {https://doi.org/10.1016/j.cpc.2021.108171} {\bibfield  {journal} {\bibinfo  {journal} {Comp. Phys. Comm.}\ }\textbf {\bibinfo {volume} {271}},\ \bibinfo {pages} {108171} (\bibinfo {year} {2022})}\BibitemShut {NoStop}%
\bibitem [{\citenamefont {Kresse}\ and\ \citenamefont {Furthm\"uller}(1996)}]{ref23VASP}%
  \BibitemOpen
  \bibfield  {author} {\bibinfo {author} {\bibfnamefont {G.}~\bibnamefont {Kresse}}\ and\ \bibinfo {author} {\bibfnamefont {J.}~\bibnamefont {Furthm\"uller}},\ }\bibfield  {title} {\bibinfo {title} {Efficient iterative schemes for ab initio total-energy calculations using a plane-wave basis set},\ }\href {https://doi.org/10.1103/PhysRevB.54.11169} {\bibfield  {journal} {\bibinfo  {journal} {Phys. Rev. B}\ }\textbf {\bibinfo {volume} {54}},\ \bibinfo {pages} {11169} (\bibinfo {year} {1996})}\BibitemShut {NoStop}%
\bibitem [{\citenamefont {Monkhorst}\ and\ \citenamefont {Pack}(1976)}]{ref25kpointsM&P}%
  \BibitemOpen
  \bibfield  {author} {\bibinfo {author} {\bibfnamefont {H.~J.}\ \bibnamefont {Monkhorst}}\ and\ \bibinfo {author} {\bibfnamefont {J.~D.}\ \bibnamefont {Pack}},\ }\bibfield  {title} {\bibinfo {title} {Special points for brillouin-zone integrations},\ }\href {https://doi.org/10.1103/PhysRevB.13.5188} {\bibfield  {journal} {\bibinfo  {journal} {Phys. Rev. B}\ }\textbf {\bibinfo {volume} {13}},\ \bibinfo {pages} {5188} (\bibinfo {year} {1976})}\BibitemShut {NoStop}%
\bibitem [{\citenamefont {Barrett}\ and\ \citenamefont {TB}(1953)}]{ref28Exp3}%
  \BibitemOpen
  \bibfield  {author} {\bibinfo {author} {\bibfnamefont {C.~S.}\ \bibnamefont {Barrett}}\ and\ \bibinfo {author} {\bibfnamefont {M.}~\bibnamefont {TB}},\ }\bibfield  {title} {\bibinfo {title} {Structure of metals: Crystallographic methods, principles, and data},\ }\href@noop {} {\bibfield  {journal} {\bibinfo  {journal} {Science}\ }\textbf {\bibinfo {volume} {117}},\ \bibinfo {pages} {421} (\bibinfo {year} {1953})}\BibitemShut {NoStop}%
\bibitem [{\citenamefont {Ihm}\ and\ \citenamefont {Cohen}(1981)}]{ref26Exp1}%
  \BibitemOpen
  \bibfield  {author} {\bibinfo {author} {\bibfnamefont {J.}~\bibnamefont {Ihm}}\ and\ \bibinfo {author} {\bibfnamefont {M.~L.}\ \bibnamefont {Cohen}},\ }\bibfield  {title} {\bibinfo {title} {Equilibrium properties and the phase transition of grey and white tin},\ }\href {https://doi.org/10.1103/PhysRevB.23.1576} {\bibfield  {journal} {\bibinfo  {journal} {Phys. Rev. B}\ }\textbf {\bibinfo {volume} {23}},\ \bibinfo {pages} {1576} (\bibinfo {year} {1981})}\BibitemShut {NoStop}%
\bibitem [{\citenamefont {Barnett}\ \emph {et~al.}(1966)\citenamefont {Barnett}, \citenamefont {Bean},\ and\ \citenamefont {Hall}}]{barnett1966x}%
  \BibitemOpen
  \bibfield  {author} {\bibinfo {author} {\bibfnamefont {J.~D.}\ \bibnamefont {Barnett}}, \bibinfo {author} {\bibfnamefont {V.~E.}\ \bibnamefont {Bean}},\ and\ \bibinfo {author} {\bibfnamefont {H.~T.}\ \bibnamefont {Hall}},\ }\bibfield  {title} {\bibinfo {title} {X-ray diffraction studies on tin to 100 kilobars},\ }\href@noop {} {\bibfield  {journal} {\bibinfo  {journal} {Journal of Applied Physics}\ }\textbf {\bibinfo {volume} {37}},\ \bibinfo {pages} {875} (\bibinfo {year} {1966})}\BibitemShut {NoStop}%
\bibitem [{\citenamefont {Bhatia}\ \emph {et~al.}(2016)\citenamefont {Bhatia}, \citenamefont {Adlakha}, \citenamefont {Lu},\ and\ \citenamefont {Solanki}}]{ref15gammaBhatia}%
  \BibitemOpen
  \bibfield  {author} {\bibinfo {author} {\bibfnamefont {M.}~\bibnamefont {Bhatia}}, \bibinfo {author} {\bibfnamefont {I.}~\bibnamefont {Adlakha}}, \bibinfo {author} {\bibfnamefont {G.}~\bibnamefont {Lu}},\ and\ \bibinfo {author} {\bibfnamefont {K.}~\bibnamefont {Solanki}},\ }\bibfield  {title} {\bibinfo {title} {Generalized stacking fault energies and slip in $\beta$-tin},\ }\href {https://doi.org/https://doi.org/10.1016/j.scriptamat.2016.05.038} {\bibfield  {journal} {\bibinfo  {journal} {Scripta Materialia}\ }\textbf {\bibinfo {volume} {123}},\ \bibinfo {pages} {21} (\bibinfo {year} {2016})}\BibitemShut {NoStop}%
\bibitem [{\citenamefont {Tang}\ \emph {et~al.}(2020)\citenamefont {Tang}, \citenamefont {Li}, \citenamefont {Deng}, \citenamefont {Wang}, \citenamefont {Zhu}, \citenamefont {Hu},\ and\ \citenamefont {Gao}}]{TANG2020103479}%
  \BibitemOpen
  \bibfield  {author} {\bibinfo {author} {\bibfnamefont {G.}~\bibnamefont {Tang}}, \bibinfo {author} {\bibfnamefont {B.}~\bibnamefont {Li}}, \bibinfo {author} {\bibfnamefont {X.}~\bibnamefont {Deng}}, \bibinfo {author} {\bibfnamefont {L.}~\bibnamefont {Wang}}, \bibinfo {author} {\bibfnamefont {W.}~\bibnamefont {Zhu}}, \bibinfo {author} {\bibfnamefont {W.}~\bibnamefont {Hu}},\ and\ \bibinfo {author} {\bibfnamefont {F.}~\bibnamefont {Gao}},\ }\bibfield  {title} {\bibinfo {title} {Comparative investigation of microjetting from tin surface subjected to laser and plane impact loadings via molecular dynamics simulations},\ }\href {https://doi.org/https://doi.org/10.1016/j.mechmat.2020.103479} {\bibfield  {journal} {\bibinfo  {journal} {Mechanics of Materials}\ }\textbf {\bibinfo {volume} {148}},\ \bibinfo {pages} {103479} (\bibinfo {year} {2020})}\BibitemShut {NoStop}%
\bibitem [{\citenamefont {Reed}\ \emph {et~al.}(2003)\citenamefont {Reed}, \citenamefont {Fried},\ and\ \citenamefont {Joannopoulos}}]{ref14MSST}%
  \BibitemOpen
  \bibfield  {author} {\bibinfo {author} {\bibfnamefont {E.~J.}\ \bibnamefont {Reed}}, \bibinfo {author} {\bibfnamefont {L.~E.}\ \bibnamefont {Fried}},\ and\ \bibinfo {author} {\bibfnamefont {J.}~\bibnamefont {Joannopoulos}},\ }\bibfield  {title} {\bibinfo {title} {A method for tractable dynamical studies of single and double shock compression},\ }\href@noop {} {\bibfield  {journal} {\bibinfo  {journal} {Physical review letters}\ }\textbf {\bibinfo {volume} {90}},\ \bibinfo {pages} {235503} (\bibinfo {year} {2003})}\BibitemShut {NoStop}%
\bibitem [{\citenamefont {Xu}\ \emph {et~al.}(2014)\citenamefont {Xu}, \citenamefont {Bi}, \citenamefont {Li}, \citenamefont {Wang}, \citenamefont {Cao}, \citenamefont {Cai}, \citenamefont {Wang},\ and\ \citenamefont {Meng}}]{ref17MPXu}%
  \BibitemOpen
  \bibfield  {author} {\bibinfo {author} {\bibfnamefont {L.}~\bibnamefont {Xu}}, \bibinfo {author} {\bibfnamefont {Y.}~\bibnamefont {Bi}}, \bibinfo {author} {\bibfnamefont {X.}~\bibnamefont {Li}}, \bibinfo {author} {\bibfnamefont {Y.}~\bibnamefont {Wang}}, \bibinfo {author} {\bibfnamefont {X.}~\bibnamefont {Cao}}, \bibinfo {author} {\bibfnamefont {L.}~\bibnamefont {Cai}}, \bibinfo {author} {\bibfnamefont {Z.}~\bibnamefont {Wang}},\ and\ \bibinfo {author} {\bibfnamefont {C.}~\bibnamefont {Meng}},\ }\bibfield  {title} {\bibinfo {title} {Phase diagram of tin determined by sound velocity measurements on multi-anvil apparatus up to 5 gpa and 800 k},\ }\bibfield  {journal} {\bibinfo  {journal} {Journal of Applied Physics}\ }\textbf {\bibinfo {volume} {115}},\ \href {https://doi.org/10.1063/1.4872458} {10.1063/1.4872458} (\bibinfo {year} {2014})\BibitemShut {NoStop}%
\bibitem [{\citenamefont {Kiefer}\ \emph {et~al.}(2002)\citenamefont {Kiefer}, \citenamefont {Duffy}, \citenamefont {Uchida},\ and\ \citenamefont {Wang}}]{kiefer2002melting}%
  \BibitemOpen
  \bibfield  {author} {\bibinfo {author} {\bibfnamefont {B.}~\bibnamefont {Kiefer}}, \bibinfo {author} {\bibfnamefont {T.}~\bibnamefont {Duffy}}, \bibinfo {author} {\bibfnamefont {T.}~\bibnamefont {Uchida}},\ and\ \bibinfo {author} {\bibfnamefont {Y.}~\bibnamefont {Wang}},\ }\bibfield  {title} {\bibinfo {title} {Melting of tin at high pressures},\ }\href@noop {} {\bibfield  {journal} {\bibinfo  {journal} {APS User Activity Report}\ } (\bibinfo {year} {2002})}\BibitemShut {NoStop}%
\bibitem [{\citenamefont {Briggs}\ \emph {et~al.}(2017)\citenamefont {Briggs}, \citenamefont {Daisenberger}, \citenamefont {Lord}, \citenamefont {Salamat}, \citenamefont {Bailey}, \citenamefont {Walter},\ and\ \citenamefont {McMillan}}]{ref16MPBriggs}%
  \BibitemOpen
  \bibfield  {author} {\bibinfo {author} {\bibfnamefont {R.}~\bibnamefont {Briggs}}, \bibinfo {author} {\bibfnamefont {D.}~\bibnamefont {Daisenberger}}, \bibinfo {author} {\bibfnamefont {O.~T.}\ \bibnamefont {Lord}}, \bibinfo {author} {\bibfnamefont {A.}~\bibnamefont {Salamat}}, \bibinfo {author} {\bibfnamefont {E.}~\bibnamefont {Bailey}}, \bibinfo {author} {\bibfnamefont {M.~J.}\ \bibnamefont {Walter}},\ and\ \bibinfo {author} {\bibfnamefont {P.~F.}\ \bibnamefont {McMillan}},\ }\bibfield  {title} {\bibinfo {title} {High-pressure melting behavior of tin up to 105 gpa},\ }\href {https://doi.org/10.1103/PhysRevB.95.054102} {\bibfield  {journal} {\bibinfo  {journal} {Phys. Rev. B}\ }\textbf {\bibinfo {volume} {95}},\ \bibinfo {pages} {054102} (\bibinfo {year} {2017})}\BibitemShut {NoStop}%
\bibitem [{\citenamefont {Fr\'eville}\ \emph {et~al.}(2024)\citenamefont {Fr\'eville}, \citenamefont {Dewaele}, \citenamefont {Guignot}, \citenamefont {Faure}, \citenamefont {Henry}, \citenamefont {Garbarino},\ and\ \citenamefont {Mezouar}}]{PhysRevB.109.104116}%
  \BibitemOpen
  \bibfield  {author} {\bibinfo {author} {\bibfnamefont {R.}~\bibnamefont {Fr\'eville}}, \bibinfo {author} {\bibfnamefont {A.}~\bibnamefont {Dewaele}}, \bibinfo {author} {\bibfnamefont {N.}~\bibnamefont {Guignot}}, \bibinfo {author} {\bibfnamefont {P.}~\bibnamefont {Faure}}, \bibinfo {author} {\bibfnamefont {L.}~\bibnamefont {Henry}}, \bibinfo {author} {\bibfnamefont {G.}~\bibnamefont {Garbarino}},\ and\ \bibinfo {author} {\bibfnamefont {M.}~\bibnamefont {Mezouar}},\ }\bibfield  {title} {\bibinfo {title} {High-pressure--high-temperature phase diagram of tin},\ }\href {https://doi.org/10.1103/PhysRevB.109.104116} {\bibfield  {journal} {\bibinfo  {journal} {Phys. Rev. B}\ }\textbf {\bibinfo {volume} {109}},\ \bibinfo {pages} {104116} (\bibinfo {year} {2024})}\BibitemShut {NoStop}%
\bibitem [{\citenamefont {Song}\ \emph {et~al.}(2016)\citenamefont {Song}, \citenamefont {Cai}, \citenamefont {Tao}, \citenamefont {Yuan}, \citenamefont {Chen}, \citenamefont {Huang}, \citenamefont {Zhao},\ and\ \citenamefont {Wang}}]{SnHugoniotMelt}%
  \BibitemOpen
  \bibfield  {author} {\bibinfo {author} {\bibfnamefont {P.}~\bibnamefont {Song}}, \bibinfo {author} {\bibfnamefont {L.-c.}\ \bibnamefont {Cai}}, \bibinfo {author} {\bibfnamefont {T.-j.}\ \bibnamefont {Tao}}, \bibinfo {author} {\bibfnamefont {S.}~\bibnamefont {Yuan}}, \bibinfo {author} {\bibfnamefont {H.}~\bibnamefont {Chen}}, \bibinfo {author} {\bibfnamefont {J.}~\bibnamefont {Huang}}, \bibinfo {author} {\bibfnamefont {X.-w.}\ \bibnamefont {Zhao}},\ and\ \bibinfo {author} {\bibfnamefont {X.-j.}\ \bibnamefont {Wang}},\ }\bibfield  {title} {\bibinfo {title} {Melting along the hugoniot and solid phase transition for sn via sound velocity measurements},\ }\href {https://doi.org/10.1063/1.4967515} {\bibfield  {journal} {\bibinfo  {journal} {Journal of Applied Physics}\ }\textbf {\bibinfo {volume} {120}},\ \bibinfo {pages} {195101} (\bibinfo {year} {2016})},\ \Eprint {https://arxiv.org/abs/https://pubs.aip.org/aip/jap/article-pdf/doi/10.1063/1.4967515/13770843/195101\_1\_online.pdf}
  {https://pubs.aip.org/aip/jap/article-pdf/doi/10.1063/1.4967515/13770843/195101\_1\_online.pdf} \BibitemShut {NoStop}%
\bibitem [{\citenamefont {Briggs}\ \emph {et~al.}(2019{\natexlab{b}})\citenamefont {Briggs}, \citenamefont {Gorman}, \citenamefont {Zhang}, \citenamefont {McGonegle}, \citenamefont {Coleman}, \citenamefont {Coppari}, \citenamefont {Morales-Silva}, \citenamefont {Smith}, \citenamefont {Wicks}, \citenamefont {Bolme}, \citenamefont {Gleason}, \citenamefont {Cunningham}, \citenamefont {Lee}, \citenamefont {Nagler}, \citenamefont {McMahon}, \citenamefont {Eggert},\ and\ \citenamefont {Fratanduono}}]{Brigs2019Coordination}%
  \BibitemOpen
  \bibfield  {author} {\bibinfo {author} {\bibfnamefont {R.}~\bibnamefont {Briggs}}, \bibinfo {author} {\bibfnamefont {M.~G.}\ \bibnamefont {Gorman}}, \bibinfo {author} {\bibfnamefont {S.}~\bibnamefont {Zhang}}, \bibinfo {author} {\bibfnamefont {D.}~\bibnamefont {McGonegle}}, \bibinfo {author} {\bibfnamefont {A.~L.}\ \bibnamefont {Coleman}}, \bibinfo {author} {\bibfnamefont {F.}~\bibnamefont {Coppari}}, \bibinfo {author} {\bibfnamefont {M.~A.}\ \bibnamefont {Morales-Silva}}, \bibinfo {author} {\bibfnamefont {R.~F.}\ \bibnamefont {Smith}}, \bibinfo {author} {\bibfnamefont {J.~K.}\ \bibnamefont {Wicks}}, \bibinfo {author} {\bibfnamefont {C.~A.}\ \bibnamefont {Bolme}}, \bibinfo {author} {\bibfnamefont {A.~E.}\ \bibnamefont {Gleason}}, \bibinfo {author} {\bibfnamefont {E.}~\bibnamefont {Cunningham}}, \bibinfo {author} {\bibfnamefont {H.~J.}\ \bibnamefont {Lee}}, \bibinfo {author} {\bibfnamefont {B.}~\bibnamefont {Nagler}}, \bibinfo {author} {\bibfnamefont {M.~I.}\ \bibnamefont {McMahon}}, \bibinfo {author}
  {\bibfnamefont {J.~H.}\ \bibnamefont {Eggert}},\ and\ \bibinfo {author} {\bibfnamefont {D.~E.}\ \bibnamefont {Fratanduono}},\ }\bibfield  {title} {\bibinfo {title} {Coordination changes in liquid tin under shock compression determined using in situ femtosecond x-ray diffraction},\ }\bibfield  {journal} {\bibinfo  {journal} {Applied Physics Letters}\ }\textbf {\bibinfo {volume} {115}},\ \href {https://doi.org/10.1063/1.5127291} {10.1063/1.5127291} (\bibinfo {year} {2019}{\natexlab{b}})\BibitemShut {NoStop}%
\bibitem [{\citenamefont {Yang}\ \emph {et~al.}(2024{\natexlab{a}})\citenamefont {Yang}, \citenamefont {Wang}, \citenamefont {Xu}, \citenamefont {Wang}, \citenamefont {Sun}, \citenamefont {Li}, \citenamefont {Zhang}, \citenamefont {Li}, \citenamefont {Yu},\ and\ \citenamefont {Wang}}]{Yang2024InsituXRD}%
  \BibitemOpen
  \bibfield  {author} {\bibinfo {author} {\bibfnamefont {J.}~\bibnamefont {Yang}}, \bibinfo {author} {\bibfnamefont {X.}~\bibnamefont {Wang}}, \bibinfo {author} {\bibfnamefont {L.}~\bibnamefont {Xu}}, \bibinfo {author} {\bibfnamefont {Q.}~\bibnamefont {Wang}}, \bibinfo {author} {\bibfnamefont {Y.}~\bibnamefont {Sun}}, \bibinfo {author} {\bibfnamefont {J.}~\bibnamefont {Li}}, \bibinfo {author} {\bibfnamefont {L.}~\bibnamefont {Zhang}}, \bibinfo {author} {\bibfnamefont {Y.}~\bibnamefont {Li}}, \bibinfo {author} {\bibfnamefont {Y.}~\bibnamefont {Yu}},\ and\ \bibinfo {author} {\bibfnamefont {P.}~\bibnamefont {Wang}},\ }\bibfield  {title} {\bibinfo {title} {Direct visualization of laser-driven dynamic fragmentation in tin by in situ x-ray diffraction},\ }\href@noop {} {\bibfield  {journal} {\bibinfo  {journal} {Matter and Radiation at Extremes}\ }\textbf {\bibinfo {volume} {9}} (\bibinfo {year} {2024}{\natexlab{a}})}\BibitemShut {NoStop}%
\bibitem [{\citenamefont {Salamat}\ \emph {et~al.}(2013)\citenamefont {Salamat}, \citenamefont {Briggs}, \citenamefont {Bouvier}, \citenamefont {Petitgirard}, \citenamefont {Dewaele}, \citenamefont {Cutler}, \citenamefont {Cor\`a}, \citenamefont {Daisenberger}, \citenamefont {Garbarino},\ and\ \citenamefont {McMillan}}]{ref19Salamat}%
  \BibitemOpen
  \bibfield  {author} {\bibinfo {author} {\bibfnamefont {A.}~\bibnamefont {Salamat}}, \bibinfo {author} {\bibfnamefont {R.}~\bibnamefont {Briggs}}, \bibinfo {author} {\bibfnamefont {P.}~\bibnamefont {Bouvier}}, \bibinfo {author} {\bibfnamefont {S.}~\bibnamefont {Petitgirard}}, \bibinfo {author} {\bibfnamefont {A.}~\bibnamefont {Dewaele}}, \bibinfo {author} {\bibfnamefont {M.~E.}\ \bibnamefont {Cutler}}, \bibinfo {author} {\bibfnamefont {F.}~\bibnamefont {Cor\`a}}, \bibinfo {author} {\bibfnamefont {D.}~\bibnamefont {Daisenberger}}, \bibinfo {author} {\bibfnamefont {G.}~\bibnamefont {Garbarino}},\ and\ \bibinfo {author} {\bibfnamefont {P.~F.}\ \bibnamefont {McMillan}},\ }\bibfield  {title} {\bibinfo {title} {High-pressure structural transformations of sn up to 138 gpa: Angle-dispersive synchrotron x-ray diffraction study},\ }\href {https://doi.org/10.1103/PhysRevB.88.104104} {\bibfield  {journal} {\bibinfo  {journal} {Phys. Rev. B}\ }\textbf {\bibinfo {volume} {88}},\ \bibinfo {pages} {104104} (\bibinfo {year}
  {2013})}\BibitemShut {NoStop}%
\bibitem [{\citenamefont {Deffrennes}\ \emph {et~al.}(2022)\citenamefont {Deffrennes}, \citenamefont {Faure}, \citenamefont {Bottin}, \citenamefont {Joubert},\ and\ \citenamefont {Oudot}}]{ref20Deffrennes}%
  \BibitemOpen
  \bibfield  {author} {\bibinfo {author} {\bibfnamefont {G.}~\bibnamefont {Deffrennes}}, \bibinfo {author} {\bibfnamefont {P.}~\bibnamefont {Faure}}, \bibinfo {author} {\bibfnamefont {F.}~\bibnamefont {Bottin}}, \bibinfo {author} {\bibfnamefont {J.-M.}\ \bibnamefont {Joubert}},\ and\ \bibinfo {author} {\bibfnamefont {B.}~\bibnamefont {Oudot}},\ }\bibfield  {title} {\bibinfo {title} {Tin (sn) at high pressure: Review, x-ray diffraction, dft calculations, and gibbs energy modeling},\ }\href {https://doi.org/https://doi.org/10.1016/j.jallcom.2022.165675} {\bibfield  {journal} {\bibinfo  {journal} {Journal of Alloys and Compounds}\ }\textbf {\bibinfo {volume} {919}},\ \bibinfo {pages} {165675} (\bibinfo {year} {2022})}\BibitemShut {NoStop}%
\bibitem [{\citenamefont {Hohenberg}\ and\ \citenamefont {Kohn}(1964{\natexlab{b}})}]{ref21DFTH&K}%
  \BibitemOpen
  \bibfield  {author} {\bibinfo {author} {\bibfnamefont {P.}~\bibnamefont {Hohenberg}}\ and\ \bibinfo {author} {\bibfnamefont {W.}~\bibnamefont {Kohn}},\ }\bibfield  {title} {\bibinfo {title} {Inhomogeneous electron gas},\ }\href {https://doi.org/10.1103/PhysRev.136.B864} {\bibfield  {journal} {\bibinfo  {journal} {Phys. Rev.}\ }\textbf {\bibinfo {volume} {136}},\ \bibinfo {pages} {B864} (\bibinfo {year} {1964}{\natexlab{b}})}\BibitemShut {NoStop}%
\bibitem [{\citenamefont {Kohn}\ and\ \citenamefont {Sham}(1965{\natexlab{b}})}]{ref22DFTK&S}%
  \BibitemOpen
  \bibfield  {author} {\bibinfo {author} {\bibfnamefont {W.}~\bibnamefont {Kohn}}\ and\ \bibinfo {author} {\bibfnamefont {L.~J.}\ \bibnamefont {Sham}},\ }\bibfield  {title} {\bibinfo {title} {Self-consistent equations including exchange and correlation effects},\ }\href {https://doi.org/10.1103/PhysRev.140.A1133} {\bibfield  {journal} {\bibinfo  {journal} {Phys. Rev.}\ }\textbf {\bibinfo {volume} {140}},\ \bibinfo {pages} {A1133} (\bibinfo {year} {1965}{\natexlab{b}})}\BibitemShut {NoStop}%
\bibitem [{\citenamefont {Aguado}(2003)}]{ref27Exp2}%
  \BibitemOpen
  \bibfield  {author} {\bibinfo {author} {\bibfnamefont {A.}~\bibnamefont {Aguado}},\ }\bibfield  {title} {\bibinfo {title} {First-principles study of elastic properties and pressure-induced phase transitions of sn: Lda versus gga results},\ }\href {https://doi.org/10.1103/PhysRevB.67.212104} {\bibfield  {journal} {\bibinfo  {journal} {Phys. Rev. B}\ }\textbf {\bibinfo {volume} {67}},\ \bibinfo {pages} {212104} (\bibinfo {year} {2003})}\BibitemShut {NoStop}%
\bibitem [{\citenamefont {Rayne}\ and\ \citenamefont {Chandrasekhar}(1960)}]{ref29Exp4}%
  \BibitemOpen
  \bibfield  {author} {\bibinfo {author} {\bibfnamefont {J.~A.}\ \bibnamefont {Rayne}}\ and\ \bibinfo {author} {\bibfnamefont {B.~S.}\ \bibnamefont {Chandrasekhar}},\ }\bibfield  {title} {\bibinfo {title} {Elastic constants of $\ensuremath{\beta}$ tin from 4.2\ifmmode^\circ\else\textdegree\fi{}k to 300\ifmmode^\circ\else\textdegree\fi{}k},\ }\href {https://doi.org/10.1103/PhysRev.120.1658} {\bibfield  {journal} {\bibinfo  {journal} {Phys. Rev.}\ }\textbf {\bibinfo {volume} {120}},\ \bibinfo {pages} {1658} (\bibinfo {year} {1960})}\BibitemShut {NoStop}%
\bibitem [{\citenamefont {Gale}\ and\ \citenamefont {Totemeier}(2003)}]{gale2003smithells}%
  \BibitemOpen
  \bibfield  {author} {\bibinfo {author} {\bibfnamefont {W.~F.}\ \bibnamefont {Gale}}\ and\ \bibinfo {author} {\bibfnamefont {T.~C.}\ \bibnamefont {Totemeier}},\ }\href@noop {} {\emph {\bibinfo {title} {Smithells metals reference book}}}\ (\bibinfo  {publisher} {Elsevier},\ \bibinfo {year} {2003})\BibitemShut {NoStop}%
\bibitem [{\citenamefont {Grigsby}\ \emph {et~al.}(2009)\citenamefont {Grigsby}, \citenamefont {Bowes}, \citenamefont {Dalton}, \citenamefont {Bernstein}, \citenamefont {Bless}, \citenamefont {Downer}, \citenamefont {Taleff}, \citenamefont {Colvin},\ and\ \citenamefont {Ditmire}}]{grigsby2009picosecond}%
  \BibitemOpen
  \bibfield  {author} {\bibinfo {author} {\bibfnamefont {W.}~\bibnamefont {Grigsby}}, \bibinfo {author} {\bibfnamefont {B.}~\bibnamefont {Bowes}}, \bibinfo {author} {\bibfnamefont {D.}~\bibnamefont {Dalton}}, \bibinfo {author} {\bibfnamefont {A.}~\bibnamefont {Bernstein}}, \bibinfo {author} {\bibfnamefont {S.}~\bibnamefont {Bless}}, \bibinfo {author} {\bibfnamefont {M.}~\bibnamefont {Downer}}, \bibinfo {author} {\bibfnamefont {E.}~\bibnamefont {Taleff}}, \bibinfo {author} {\bibfnamefont {J.}~\bibnamefont {Colvin}},\ and\ \bibinfo {author} {\bibfnamefont {T.}~\bibnamefont {Ditmire}},\ }\bibfield  {title} {\bibinfo {title} {Picosecond time scale dynamics of short pulse laser-driven shocks in tin},\ }\href@noop {} {\bibfield  {journal} {\bibinfo  {journal} {Journal of Applied Physics}\ }\textbf {\bibinfo {volume} {105}} (\bibinfo {year} {2009})}\BibitemShut {NoStop}%
\bibitem [{\citenamefont {La~Lone}\ \emph {et~al.}(2013)\citenamefont {La~Lone}, \citenamefont {Stevens}, \citenamefont {Turley}, \citenamefont {Holtkamp}, \citenamefont {Iverson}, \citenamefont {Hixson},\ and\ \citenamefont {Veeser}}]{la2013release}%
  \BibitemOpen
  \bibfield  {author} {\bibinfo {author} {\bibfnamefont {B.}~\bibnamefont {La~Lone}}, \bibinfo {author} {\bibfnamefont {G.}~\bibnamefont {Stevens}}, \bibinfo {author} {\bibfnamefont {W.}~\bibnamefont {Turley}}, \bibinfo {author} {\bibfnamefont {D.}~\bibnamefont {Holtkamp}}, \bibinfo {author} {\bibfnamefont {A.}~\bibnamefont {Iverson}}, \bibinfo {author} {\bibfnamefont {R.}~\bibnamefont {Hixson}},\ and\ \bibinfo {author} {\bibfnamefont {L.}~\bibnamefont {Veeser}},\ }\bibfield  {title} {\bibinfo {title} {Release path temperatures of shock-compressed tin from dynamic reflectance and radiance measurements},\ }\href@noop {} {\bibfield  {journal} {\bibinfo  {journal} {Journal of Applied Physics}\ }\textbf {\bibinfo {volume} {114}} (\bibinfo {year} {2013})}\BibitemShut {NoStop}%
\bibitem [{\citenamefont {Signor}\ \emph {et~al.}(2010)\citenamefont {Signor}, \citenamefont {de~Ress{\'e}guier}, \citenamefont {Dragon}, \citenamefont {Roy}, \citenamefont {Fanget},\ and\ \citenamefont {Faessel}}]{signor2010investigation}%
  \BibitemOpen
  \bibfield  {author} {\bibinfo {author} {\bibfnamefont {L.}~\bibnamefont {Signor}}, \bibinfo {author} {\bibfnamefont {T.}~\bibnamefont {de~Ress{\'e}guier}}, \bibinfo {author} {\bibfnamefont {A.}~\bibnamefont {Dragon}}, \bibinfo {author} {\bibfnamefont {G.}~\bibnamefont {Roy}}, \bibinfo {author} {\bibfnamefont {A.}~\bibnamefont {Fanget}},\ and\ \bibinfo {author} {\bibfnamefont {M.}~\bibnamefont {Faessel}},\ }\bibfield  {title} {\bibinfo {title} {Investigation of fragments size resulting from dynamic fragmentation in melted state of laser shock-loaded tin},\ }\href@noop {} {\bibfield  {journal} {\bibinfo  {journal} {International Journal of Impact Engineering}\ }\textbf {\bibinfo {volume} {37}},\ \bibinfo {pages} {887} (\bibinfo {year} {2010})}\BibitemShut {NoStop}%
\bibitem [{\citenamefont {Sorenson}\ \emph {et~al.}(2002)\citenamefont {Sorenson}, \citenamefont {Minich}, \citenamefont {Romero}, \citenamefont {Tunnell},\ and\ \citenamefont {Malone}}]{sorenson2002ejecta}%
  \BibitemOpen
  \bibfield  {author} {\bibinfo {author} {\bibfnamefont {D.}~\bibnamefont {Sorenson}}, \bibinfo {author} {\bibfnamefont {R.}~\bibnamefont {Minich}}, \bibinfo {author} {\bibfnamefont {J.}~\bibnamefont {Romero}}, \bibinfo {author} {\bibfnamefont {T.}~\bibnamefont {Tunnell}},\ and\ \bibinfo {author} {\bibfnamefont {R.}~\bibnamefont {Malone}},\ }\bibfield  {title} {\bibinfo {title} {Ejecta particle size distributions for shock loaded sn and al metals},\ }\href@noop {} {\bibfield  {journal} {\bibinfo  {journal} {Journal of applied physics}\ }\textbf {\bibinfo {volume} {92}},\ \bibinfo {pages} {5830} (\bibinfo {year} {2002})}\BibitemShut {NoStop}%
\bibitem [{\citenamefont {Katzke}\ \emph {et~al.}(2006)\citenamefont {Katzke}, \citenamefont {Bismayer},\ and\ \citenamefont {Tol\'edano}}]{PhysRevB.73.134105}%
  \BibitemOpen
  \bibfield  {author} {\bibinfo {author} {\bibfnamefont {H.}~\bibnamefont {Katzke}}, \bibinfo {author} {\bibfnamefont {U.}~\bibnamefont {Bismayer}},\ and\ \bibinfo {author} {\bibfnamefont {P.}~\bibnamefont {Tol\'edano}},\ }\bibfield  {title} {\bibinfo {title} {Theory of the high-pressure structural phase transitions in si, ge, sn, and pb},\ }\href {https://doi.org/10.1103/PhysRevB.73.134105} {\bibfield  {journal} {\bibinfo  {journal} {Phys. Rev. B}\ }\textbf {\bibinfo {volume} {73}},\ \bibinfo {pages} {134105} (\bibinfo {year} {2006})}\BibitemShut {NoStop}%
\bibitem [{\citenamefont {Brouillette}(2002)}]{RMI}%
  \BibitemOpen
  \bibfield  {author} {\bibinfo {author} {\bibfnamefont {M.}~\bibnamefont {Brouillette}},\ }\bibfield  {title} {\bibinfo {title} {The richtmyer-meshkov instability},\ }\href {https://doi.org/https://doi.org/10.1146/annurev.fluid.34.090101.162238} {\bibfield  {journal} {\bibinfo  {journal} {Annual Review of Fluid Mechanics}\ }\textbf {\bibinfo {volume} {34}},\ \bibinfo {pages} {445} (\bibinfo {year} {2002})}\BibitemShut {NoStop}%
\bibitem [{\citenamefont {Grady}(1988)}]{GRADY1988353}%
  \BibitemOpen
  \bibfield  {author} {\bibinfo {author} {\bibfnamefont {D.}~\bibnamefont {Grady}},\ }\bibfield  {title} {\bibinfo {title} {The spall strength of condensed matter},\ }\href {https://doi.org/https://doi.org/10.1016/0022-5096(88)90015-4} {\bibfield  {journal} {\bibinfo  {journal} {Journal of the Mechanics and Physics of Solids}\ }\textbf {\bibinfo {volume} {36}},\ \bibinfo {pages} {353} (\bibinfo {year} {1988})}\BibitemShut {NoStop}%
\bibitem [{\citenamefont {Saunders}\ \emph {et~al.}(2021)\citenamefont {Saunders}, \citenamefont {Stan}, \citenamefont {Mackay}, \citenamefont {Morgan}, \citenamefont {Horwitz}, \citenamefont {Ali}, \citenamefont {Rinderknecht}, \citenamefont {Haxhimali}, \citenamefont {Ping}, \citenamefont {Najjar}, \citenamefont {Eggert},\ and\ \citenamefont {Park}}]{PhysRevLett.127.155002}%
  \BibitemOpen
  \bibfield  {author} {\bibinfo {author} {\bibfnamefont {A.~M.}\ \bibnamefont {Saunders}}, \bibinfo {author} {\bibfnamefont {C.~V.}\ \bibnamefont {Stan}}, \bibinfo {author} {\bibfnamefont {K.~K.}\ \bibnamefont {Mackay}}, \bibinfo {author} {\bibfnamefont {B.}~\bibnamefont {Morgan}}, \bibinfo {author} {\bibfnamefont {J.~A.~K.}\ \bibnamefont {Horwitz}}, \bibinfo {author} {\bibfnamefont {S.~J.}\ \bibnamefont {Ali}}, \bibinfo {author} {\bibfnamefont {H.~G.}\ \bibnamefont {Rinderknecht}}, \bibinfo {author} {\bibfnamefont {T.}~\bibnamefont {Haxhimali}}, \bibinfo {author} {\bibfnamefont {Y.}~\bibnamefont {Ping}}, \bibinfo {author} {\bibfnamefont {F.}~\bibnamefont {Najjar}}, \bibinfo {author} {\bibfnamefont {J.}~\bibnamefont {Eggert}},\ and\ \bibinfo {author} {\bibfnamefont {H.-S.}\ \bibnamefont {Park}},\ }\bibfield  {title} {\bibinfo {title} {Experimental observations of laser-driven tin ejecta microjet interactions},\ }\href {https://doi.org/10.1103/PhysRevLett.127.155002} {\bibfield  {journal} {\bibinfo  {journal}
  {Phys. Rev. Lett.}\ }\textbf {\bibinfo {volume} {127}},\ \bibinfo {pages} {155002} (\bibinfo {year} {2021})}\BibitemShut {NoStop}%
\bibitem [{\citenamefont {Yang}\ \emph {et~al.}(2024{\natexlab{b}})\citenamefont {Yang}, \citenamefont {Wang}, \citenamefont {Xu}, \citenamefont {Wang}, \citenamefont {Sun}, \citenamefont {Li}, \citenamefont {Zhang}, \citenamefont {Li}, \citenamefont {Yu}, \citenamefont {Wang}, \citenamefont {Wu},\ and\ \citenamefont {Hu}}]{Yang2024MRE}%
  \BibitemOpen
  \bibfield  {author} {\bibinfo {author} {\bibfnamefont {J.}~\bibnamefont {Yang}}, \bibinfo {author} {\bibfnamefont {X.}~\bibnamefont {Wang}}, \bibinfo {author} {\bibfnamefont {L.}~\bibnamefont {Xu}}, \bibinfo {author} {\bibfnamefont {Q.}~\bibnamefont {Wang}}, \bibinfo {author} {\bibfnamefont {Y.}~\bibnamefont {Sun}}, \bibinfo {author} {\bibfnamefont {J.}~\bibnamefont {Li}}, \bibinfo {author} {\bibfnamefont {L.}~\bibnamefont {Zhang}}, \bibinfo {author} {\bibfnamefont {Y.}~\bibnamefont {Li}}, \bibinfo {author} {\bibfnamefont {Y.}~\bibnamefont {Yu}}, \bibinfo {author} {\bibfnamefont {P.}~\bibnamefont {Wang}}, \bibinfo {author} {\bibfnamefont {Q.}~\bibnamefont {Wu}},\ and\ \bibinfo {author} {\bibfnamefont {J.}~\bibnamefont {Hu}},\ }\bibfield  {title} {\bibinfo {title} {Direct visualization of laser-driven dynamic fragmentation in tin by in situ x-ray diffraction},\ }\href {https://doi.org/10.1063/5.0200242} {\bibfield  {journal} {\bibinfo  {journal} {Matter and Radiation at Extremes}\ }\textbf {\bibinfo {volume}
  {9}},\ \bibinfo {pages} {057803} (\bibinfo {year} {2024}{\natexlab{b}})}\BibitemShut {NoStop}%
\bibitem [{\citenamefont {Zaretsky}\ and\ \citenamefont {Kanel}(2012)}]{JAP2012Al}%
  \BibitemOpen
  \bibfield  {author} {\bibinfo {author} {\bibfnamefont {E.~B.}\ \bibnamefont {Zaretsky}}\ and\ \bibinfo {author} {\bibfnamefont {G.~I.}\ \bibnamefont {Kanel}},\ }\bibfield  {title} {\bibinfo {title} {Effect of temperature, strain, and strain rate on the flow stress of aluminum under shock-wave compression},\ }\href {https://doi.org/10.1063/1.4755792} {\bibfield  {journal} {\bibinfo  {journal} {Journal of Applied Physics}\ }\textbf {\bibinfo {volume} {112}},\ \bibinfo {pages} {073504} (\bibinfo {year} {2012})}\BibitemShut {NoStop}%
\bibitem [{\citenamefont {Chen}\ \emph {et~al.}(2016)\citenamefont {Chen}, \citenamefont {Ren}, \citenamefont {Tang}, \citenamefont {Li},\ and\ \citenamefont {Hu}}]{ShockWaves2016}%
  \BibitemOpen
  \bibfield  {author} {\bibinfo {author} {\bibfnamefont {Y.}~\bibnamefont {Chen}}, \bibinfo {author} {\bibfnamefont {G.}~\bibnamefont {Ren}}, \bibinfo {author} {\bibfnamefont {T.}~\bibnamefont {Tang}}, \bibinfo {author} {\bibfnamefont {Q.}~\bibnamefont {Li}},\ and\ \bibinfo {author} {\bibfnamefont {H.}~\bibnamefont {Hu}},\ }\bibfield  {title} {\bibinfo {title} {Experimental study of micro-spalling fragmentation from melted lead},\ }\href {https://doi.org/10.1007/s00193-015-0601-4} {\bibfield  {journal} {\bibinfo  {journal} {Shock Waves}\ }\textbf {\bibinfo {volume} {26}},\ \bibinfo {pages} {221} (\bibinfo {year} {2016})}\BibitemShut {NoStop}%
\bibitem [{\citenamefont {Wen}\ \emph {et~al.}(2022)\citenamefont {Wen}, \citenamefont {Zhang}, \citenamefont {Wang}, \citenamefont {E},\ and\ \citenamefont {Srolovitz}}]{Wen_2022}%
  \BibitemOpen
  \bibfield  {author} {\bibinfo {author} {\bibfnamefont {T.}~\bibnamefont {Wen}}, \bibinfo {author} {\bibfnamefont {L.}~\bibnamefont {Zhang}}, \bibinfo {author} {\bibfnamefont {H.}~\bibnamefont {Wang}}, \bibinfo {author} {\bibfnamefont {W.}~\bibnamefont {E}},\ and\ \bibinfo {author} {\bibfnamefont {D.~J.}\ \bibnamefont {Srolovitz}},\ }\bibfield  {title} {\bibinfo {title} {Deep potentials for materials science},\ }\href {https://doi.org/10.1088/2752-5724/ac681d} {\bibfield  {journal} {\bibinfo  {journal} {Materials Futures}\ }\textbf {\bibinfo {volume} {1}},\ \bibinfo {pages} {022601} (\bibinfo {year} {2022})}\BibitemShut {NoStop}%
\end{thebibliography}%

\end{document}